\documentclass[10pt]{IEEEtran}

\usepackage{graphicx}          
\usepackage{graphics} 
\usepackage{epsfig} 
\usepackage{mathptmx} 
\usepackage{times} 
\usepackage{amsmath} 
\usepackage{amssymb}  
\usepackage{ifthen}
\usepackage{booktabs}
\usepackage{algorithm}
\usepackage{algorithmic}
\usepackage{setspace}
\usepackage{amsmath}
\usepackage{amscd}
\newtheorem{theorem}{Theorem}
\newtheorem{lemma}{Lemma}
\newtheorem{remark}{Remark}
\begin{document}
%
\title{Event-Triggered State Estimation Through Confidence Level}

\author{Wei Liu, \textit{Senior Member}, \textit{IEEE}

\thanks{This work was partially supported by the National Nature Science Foundation of China (6207312).}

\thanks{Wei Liu is with the School of Information and Electronic Engineering, Zhejiang Gongshang University, Hangzhou 310018,
China (e-mail: intervalm@163.com).}
}

\markboth{}%
{Shell \MakeLowercase{\textit{et al.}}: Bare Demo of IEEEtran.cls for Journals}

\maketitle

\vspace{-25pt}
\begin{abstract}
This paper considers the state estimation problem for discrete-time linear systems under event-triggered scheme.
In order to improve performance, a novel event-triggered scheme based on confidence level is proposed using the chi-square
distribution and mild regularity assumption. In terms of the novel event-triggered scheme, a minimum mean squared error (MMSE) state estimator is proposed
using some results presented in this paper. Two algorithms for communication rate estimation of the proposed MMSE state estimator are developed where the
first algorithm is based on information with one-step delay, and the second algorithm is based on information with two-step delay. The performance and effectiveness of the proposed MMSE state estimator
and the two communication rate estimation algorithms are illustrated
using a target tracking scenario.
\end{abstract}

\vspace{-5pt}
\section{Introduction}
\label{}
With the development of wireless sensor network technology, wireless networked control systems (WNCS) have attracted increasing attention, and have been successfully applied in various fields
such as control, signal processing, robotics, power electronics, etc \cite{ap1}$-$\cite{ap6}.
In WNCS, sensors, controllers, estimators and actuators are spatially distributed where sensor and estimator are usually far away from each other.
In this case, the communication from sensor to remote estimator is costly because the communication requires consuming power of energy limited
battery in the sensor where the battery is probably hard to replace due to its physical position. Event-triggered scheme is an effective means to reduce
sensor-to-estimator communication cost since communication is not permitted unless a pre-defined triggered condition
is satisfied.
Previous studies have shown that event-triggered scheme can strike a proper balance of trade-offs between communication cost and estimation performance \cite{rewCr}$-$\cite{et1}.
\par
As a fundamental issue, event-triggered state estimation has been extensively studied \cite{et1}$-$\cite{etad2}.
In \cite{et2}, for a first order discrete-time linear system, the pre-processor
and estimator were sought to minimize a cost with two terms.
In \cite{et3},
a centralized sensor network with multiple nodes were considered where each node yields measurement of the original system.
Local event-triggered transmission strategies were developed, and the strategies's stability and performance were studied.
For the balance between communication rate and state estimation performance, an event-triggered sensor data
scheduler was presented in
\cite{et1} where, for a specific threshold, this scheduler is determined by the H\"{o}lder infinity-norm of the
innovation's linear operation. Using an approximation technique in nonlinear filtering, an approximate MMSE state estimator
was proposed. The results of \cite{et1} were extended in \cite{et1a} and \cite{et5} where the results presented in \cite{et1a} considered
separate transmission for each element of the measurement, and measurements from multiple sensors with separate event-triggering conditions for each
measurement were studied in \cite{et5}. The measurement
prediction variance was used in \cite{et6} to determine whether the measurement is transmitted. Based on this kind of measurement transmission,
the state estimator was designed, and the corresponding Riccati
equation with periodic behavior was developed.
In \cite{et7}, the set-valued Kalman filtering problem for additional information with stochastic
uncertainty was studied, and it was applied to event-triggered estimation.
Two stochastic
event-triggered sensor schedules were proposed in \cite{et8} where one schedule depends on the current measurement, and the other one
depends on the innovation. Based on the two schedules, the MMSE state estimators were proposed, and
the communication rates were analyzed. The results of \cite{et8} were generalized and extended in \cite{et9} and \cite{et10}, respectively, where
single-sensor was generalized to the case of multi-sensor in \cite{et9}, and a stochastic
event-triggered mechanism based on information-state contribution was proposed in \cite{et10}. More results about event-triggered state estimation
were provided in  \cite{et11}$-$\cite{etad2} and references therein. Because additional information was introduced to the remote estimator when the measurement
is not transmitted to the estimator, the state estimator developed in \cite{et1} can yield better performance. However, the results developed in \cite{et1}
does not establish the connection between the innovation and the trigger threshold, which means the performance can be further improved through establishing a
proper connection between them.  So, it is necessary to propose an event-triggered scheme which can establish a proper connection of
the innovation, the trigger threshold and other related parameters, and to design state estimator based on this scheme, which motivates our research.
\par
In this paper, using the chi-square
distribution, regular Gaussian assumption and the method of confidence level,  we first propose an event-triggered scheme which establishes a proper connection of
the tolerable upper bound of the innovation covariance,
the
innovation and the trigger threshold. However, the results proposed in \cite{et1} do not obtain any connection of these parameters. Also, to the best of the author's knowledge,  the
event-triggered scheme proposed in this paper is novel and different from the existing results. The novel event-triggered scheme paves the way for the design of state estimator with
better performance.
\par
Then, based on the novel event-triggered scheme, a MMSE state estimator is proposed in a recursive form. It is worth mentioning that, due to the use of the novel
event-triggered scheme,
 the strategy in computing the error covariance of the proposed MMSE state estimator is different in contrast to the existing results. Two algorithms
 for estimating the communication rate of the proposed state estimator are developed where the first algorithm uses information with one-step delay, and the second algorithm
 utilizes information with two-step delay. As far as the author knows, the strategy used in the second algorithm, namely, using information with two-step delay,
cannot be found in the existing results for estimating the communication rate of event-triggered
state estimator. In addition, the simulation results show that the second algorithm yields a better communication rate estimation of the proposed state estimator, which means
that the strategy of using information with two-step delay is effective. Due to using information with two-step delay, the proof of the second algorithm becomes very challenging.
In order to prove the second algorithm, we first prove Lemma \ref{lemyeqz} and Theorem \ref{Theorem2} in Appendix \ref{appenT4}, and then we prove the second
algorithm in Appendix \ref{appenT5}.
\par
The remainder of this paper is organized as follows. The system and problem under consideration are provided in Section \ref{sec2}.
We also present an
event-triggered scheme in Section \ref{sec2}.
A  MMSE state estimator based on the presented event-triggered scheme is proposed in Section \ref{secmro}.
Two algorithms for estimating the communication rate of the proposed  MMSE state estimator are developed in Section
\ref{sbd}. In Section  \ref{sectionNE}, the performance  and effectiveness  of the proposed results, including the MMSE state estimator and the communication rate estimation algorithms, are demonstrated via a target tracking scenario.
The conclusion is drawn in Section \ref{sectionCo}.
\par
    \textit{Notation:}
    The $n$-dimensional real Euclidean space is denoted by
$\mathbb{R}^n$, and $N>0$ is used to denote the positive definite matrix $N$.
For a matrix $A$, its transpose, determinant and inverse are represented by $A^\textrm{T}$, $|A|$ and $A^{-1}$, respectively.
The probability density function is denoted by $f$, and the $n\times n$ identity matrix is denoted by $\textsl{\textsf{I}}_n$.
$\int\varPhi(\eta)d\eta$ is used to stand for $\int\varPhi(\eta)d\eta_1d\eta_2\cdots d\eta_n$ where $\eta=(\eta_1,\eta_2,\cdots,\eta_n)^\textrm{T}\in\mathbb{R}^n$, and $\varPhi(\eta)$ is a function of $\eta$.
We use $\textrm{E}[\cdot]$ and $\textrm{Var}(\cdot)$ to stand for the expectation operation and the covariance operation, respectively.
\section{Problem Formulation  \label{sec2}}
\subsection{ System Description\label{sbecmesy}}  Consider the following system
\begin{align}x_{k+1}=&Ax_{k}+\omega_{k},\label{sxk}\\
y_{k}=&Cx_k+\upsilon_{k}, k=0,1,\cdots\label{syk}
\end{align} where \(x_{k}\in\mathbb{R}^n\) is
the unknown state;
 \(\omega_k\in\mathbb{R}^n\) is the process noise; \(y_{k}\in\mathbb{R}^{p}\) is the measurement;  \(\upsilon_{k}\in\mathbb{R}^{p}\)
 is the measurement noise; \(A\) and $C$ are matrices of appropriate dimensions; and
 the initial state \(x_0\) is a random
vector with mean $\bar{x}_0$ and covariance matrix \(\bar{P}_0\).
\par
Throughout the paper, we introduce the following two assumptions.
\begin{enumerate}
  \item $\omega_{k}$ and $\upsilon_{k}$ are zero-mean white Gaussian noise sequences with covariance
matrices $Q$ and $R$, respectively.
  \item $\omega_{k}$ is independent of $\upsilon_{k}$, and
 $x_0$ is independent of $\omega_k$ and $\upsilon_{k}$.
\end{enumerate}
\par
\begin{figure}
\vspace{4pt}
  \begin{center}
\includegraphics[ width=3.4in]{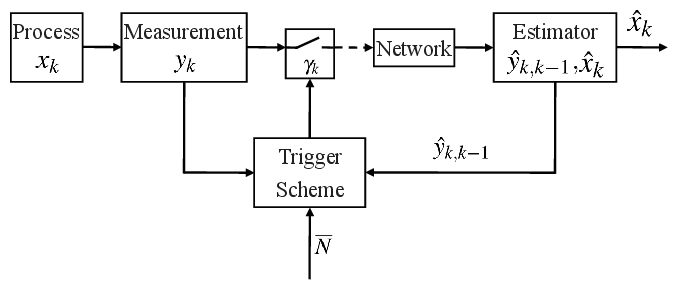}
\caption{  Structure of event-triggered state estimation.}
  \label{figst}
  \end{center}
  \vspace{-15pt}
\end{figure}
Considering the event-triggered state estimation problem whose structure is given in Fig. \ref{figst}.
$\gamma_{k}$ has two possible values 0 and 1, and the value of $\gamma_{k}$ is determined by the trigger
scheme. When $\gamma_{k}=1$, the measurement $y_k$ is transmitted to the estimator via network, and the information $\textrm{I}_k$ available for the estimator is
$\textrm{I}_k=(\textrm{I}_{k-1},y_{k}).$ When $\gamma_{k}=0$, there is no data tramission, and the information $\textrm{I}_k$ available for the estimator is
$\textrm{I}_{k}=(\textrm{I}_{k-1},\gamma_{k}=0)$. Hence,\vspace{3pt} we have
$\textrm{I}_k=\left\{
                                                                                        \begin{array}{ll}
                                                                                          (\textrm{I}_{k-1},\gamma_{k}=0), & \gamma_{k}=0; \\
                                                                                          (\textrm{I}_{k-1},y_k), & \gamma_{k}=1
                                                                                        \end{array}
                                                                                      \right.$\vspace{3pt}  with $\textrm{I}_{0}=\left\{
                                                                                        \begin{array}{ll}
                                                                                          (\gamma_{0}=0), & \gamma_{0}=0; \\
                                                                                          (y_0), & \gamma_{0}=1
                                                                                        \end{array}
                                                                                      \right.$.
%
The trigger
scheme is based on confidence level, and it will be proposed in the next part.
\begin{remark}
For a joint probability density function $f(x,y)$ in probability theory, $f(x=a,y=b)$ denotes the joint probability density of $x$ and $y$
given $(x=a,y=b)$ where $(x=a,y=b)$ stands for $\{x=a\}\cap\{ y=b\}$ instead of $\{x=a\}\cup\{ y=b\}$. As an extension,
the conditional probability density of $x=a$ given $y=b$ is defined as $f(x=a|y=b)=\frac{f(x=a,y=b)}{f(y=b)}$. It has to be
said that the conditional probability $f(x=a|y=b)$ is not equal to $\frac{f(x=a \cup y=b)}{f(y=b)}$, that is, $f(x=a|y=b)\neq \frac{f(x=a \cup y=b)}{f(y=b)}$,  where $(x=a \cup y=b)$ denotes  $\{x=a\}\cup\{ y=b\}$ which is the union of two random variables.
The conditional probability was used in the wrong way in \cite{et1}, which expressed the probability density with the union of two random variables. For example, in \cite{et1},
the term $f_{\epsilon_k}(\epsilon|I_{k-1})$ is equal to $\frac{f(\epsilon_k=\epsilon \cup I_{k-1})}{f(I_{k-1})}$, and $f_{\epsilon_k}(\epsilon|I_{k-1})$ is obviously not equal to
the conditional probability density of $\epsilon_k=\epsilon$ given $I_{k-1}$. However, the term $f_{\epsilon_k}(\epsilon|I_{k-1})$ was used as the conditional probability density,
that is, the results were developed based on the condition $\frac{f(\epsilon_k=\epsilon \cup I_{k-1})}{f(I_{k-1})}=\frac{f(\epsilon_k=\epsilon, I_{k-1})}{f(I_{k-1})}$.
Hence,
the correctness of the corresponding results presented in \cite{et1} is doubtful unless the information $I_{k-1}\cap \{\gamma_k=0\}$ is used to replace $I_{k-1}\cup \{\gamma_k=0\}$.
%
%
%
%
\label{Remarkep}
\end{remark}
\begin{remark}
In \cite{et8},
${I}_k$ is defined by ${I}_k\triangleq \{\gamma_0,\gamma_1,\cdots,\gamma_k,\cdots,\gamma_0y_0,\gamma_1y_1,\gamma_ky_k\}=\{{I}_{k-1},\gamma_k,\gamma_ky_k\}$.
Because the relation between ${I}_{k-1}$, $\gamma_k$ and $\gamma_ky_k$ in ${I}_k$ is not clearly stated,\vspace{2pt} we first assume that
${I}_k={I}_{k-1}\cap (\gamma_k\cup \gamma_ky_k)$. Then, we have $\gamma_k\cup \gamma_ky_k=\left\{
                                                                                        \begin{array}{ll}
                                                                                          \gamma_{k}=0, & \gamma_{k}=0; \\
                                                                                          (\gamma_{k}=1)\cup y_k, & \gamma_{k}=1
                                                                                        \end{array}
                                                                                      \right.$, which means that
${I}_k=\left\{
                                                                                        \begin{array}{ll}
                                                                                          {I}_{k-1}\cap (\gamma_{k}=0), & \gamma_{k}=0; \\
                                                                                          {I}_{k-1}\cap \big((\gamma_{k}=1)\cup y_k\big), & \gamma_{k}=1
                                                                                        \end{array}
                                                                                      \right.$. We easily see that, under the above assumption, ${I}_k$ is equal to
the description presented in \cite{et8} for $\gamma_{k}=0$, but is not equal to
that for $\gamma_{k}=1$. Second, we assume that ${I}_k={I}_{k-1}\cap (\gamma_k\cap \gamma_ky_k)$. Then, we get $\gamma_k\cap \gamma_ky_k=0$ in $\gamma_{k}=0$.
As a result, we derive ${I}_k={I}_{k-1}$  in $\gamma_{k}=0$. It is obvious that, under the assumption of ${I}_k={I}_{k-1}\cap (\gamma_k\cap \gamma_ky_k)$, ${I}_k$ is not equal to
the description presented in \cite{et8} for $\gamma_{k}=0$. Hence,  no matter how we choose the relation between ${I}_{k-1}$, $\gamma_k$ and $\gamma_ky_k$, ${I}_k$\vspace{2pt} cannot completely embody the description
${I}_k=\left\{
                                                                                        \begin{array}{ll}
                                                                                         (\textrm{I}_{k-1},\gamma_{k}=0), & \gamma_{k}=0; \\
                                                                                          ({I}_{k-1},y_k), & \gamma_{k}=1
                                                                                        \end{array}
                                                                                      \right.$ presented in \cite{et8}. This means that the definition of
 ${I}_k\triangleq \{\gamma_0,\gamma_1,\cdots,\gamma_k,\cdots,\gamma_0y_0,\gamma_1y_1,\gamma_ky_k\}$ given in \cite{et8} is not rigorous.
\label{Remarksp}
\end{remark}
\subsection{ Event-Triggered Scheme based on Confidence Level\label{sbecmev}}
In order to propose the event-triggered scheme based on confidence level, we first present the following two remarks and one lemma.
\begin{remark}
Let $w\in\mathbb{R}^{p}$ be a Gaussian random vector with mean $\bar{w}$ and covariance matrix $S$. Then, it was presented in Result 4.7 of \cite{b1} that
$(w-\bar{w})^\textrm{T}S^{-1}(w-\bar{w})$ obeys the distribution of $\chi_p^2$  where $\chi_p^2$ stands for  the chi-square
distribution with $p$ degrees of freedom.
\label{Remarkeg1}
\end{remark}
\begin{remark}
Let $\mu_1$, $\mu_2$, $\cdots$, $\mu_m$ be Gaussian and mutually independent. Then, it is well known that the linear combination of  $\mu_1$, $\mu_2$, $\cdots$, $\mu_m$ is still Gaussian.
\label{Remarkeig}
\end{remark}
\begin{lemma}
Under the assumption that $f(x_{k-1}|\textrm{I}_{k-1})$ is Gaussian, it holds that:\\
1). $f(x_{k}|\textrm{I}_{k-1})$ is Gaussian.\\
2). $f(y_{k}|\textrm{I}_{k-1})$ is Gaussian.
\label{lemfGh}
\end{lemma}
\par
\textit{Proof:} See Appendix \ref{appenT1}.
\par
For notational simplicity, define
\begin{align}
\hat{y}_{k,k-1}\triangleq& \textrm{E}[y_k|\textrm{I}_{k-1}],\label{nota}\\
 \tilde{y}_k\triangleq& y_k-\hat{y}_{k,k-1},\label{notb}\\
 N_k\triangleq&\textrm{Var}(y_k|\textrm{I}_{k-1}).\label{notb1a}
\end{align}
\par
Under the assumption that $f(x_{k-1}|\textrm{I}_{k-1})$ is Gaussian, we see from 2) of Lemma \ref{lemfGh} that $f(y_{k}|\textrm{I}_{k-1})$ is Gaussian.
Then, using Remark \ref{Remarkeg1}, we find that  $\tilde{y}_k^\textrm{T}N_k^{-1}\tilde{y}_k$ is distributed as  $\chi_p^2$ where
$\tilde{y}_k$ and $N_k$ are defined in (\ref{notb}) and (\ref{notb1a}), respectively.
Let $\overline{N}>0$ be a tolerable upper bound of $N_k$.  When $N_k$ exceeds the tolerable upper bound, namely,  $N_k>\overline{N}$, we need to
take $\gamma_k=1$ so that $N_{k+1}$ does not exceed the tolerable upper bound. When  $N_k>\overline{N}$, we have
\begin{align}
\varphi_k>\tilde{y}_k^\textrm{T}N_k^{-1}\tilde{y}_k
\label{eva}
\end{align}
where
\begin{align}
\varphi_k=\tilde{y}_k^\textrm{T}\varSigma\tilde{y}_k, \varSigma=\overline{N}^{-1}.\label{ySy}
\end{align}
Let $\chi_\alpha^2(p)$ denote the upper (l00$\alpha$)th  percentile of the $\chi_p^2$ distribution, that is, $P\big(\tilde{y}_k^\textrm{T}N_k^{-1}\tilde{y}_k\leq\chi_\alpha^2(p)\big)=1-\alpha$.
In frequentist statistics, the 95\% confidence level is the most often. Hence, we can take $1-\alpha=0.95$ in confidence level for $\tilde{y}_k^\textrm{T}N_k^{-1}\tilde{y}_k$.
Then, we conclude from the theory of confidence level that $\tilde{y}_k^\textrm{T}N_k^{-1}\tilde{y}_k$ does not obey the $\chi_p^2$ distribution if $\tilde{y}_k^\textrm{T}N_k^{-1}\tilde{y}_k>\chi_\alpha^2(p)$. Then,\vspace{2pt}
using (\ref{eva}), we get $\varphi_k>\tilde{y}_k^\textrm{T}N_k^{-1}\tilde{y}_k>\chi_\alpha^2(p)\Longrightarrow \varphi_k>\chi_\alpha^2(p)$ when $N_k>\overline{N}$. Based on the above discussion, we present the following
event-triggered scheme:
\begin{align}
\gamma_{k}=&\begin{cases}
\gamma_{k}=0,&\varphi_k\leq \chi_\alpha^2(p);
\\
\gamma_{k}=1,&\varphi_k> \chi_\alpha^2(p)
.\label{et}\\
\end{cases}
\end{align}
where $1-\alpha=0.95$ is suggested to be taken considering the theory of confidence level.
Since $\varSigma$ is positive definite, $\varSigma$ can be expressed as $\varSigma=\varPhi^\textrm{T}\varPhi$ such that $\varPhi$ is invertible.
Then, using (\ref{ySy}) and noticing \(y_{k}\in\mathbb{R}^{p}\), we have
\begin{align}
\varphi_k=\tilde{y}_k^\textrm{T}\varSigma\tilde{y}_k=\tilde{y}_k^\textrm{T}\varPhi^\textrm{T}\varPhi\tilde{y}_k=z_k^\textrm{T}z_k=\sum_{i=1}^pz_{k,i}^2\label{ySya}
\end{align}
where
\begin{align}
z_k\triangleq \varPhi\tilde{y}_k\label{zk}
\end{align}
and $z_{k,i}$ denotes the $i$th element of $z_k$.
\begin{remark} An advantage of the event-triggered scheme proposed in this paper is that a proper connection of the tolerable upper bound $\overline{N}$,
the
innovation $\tilde{y}_k$ and the trigger threshold $\chi_\alpha^2(p)$ is established via confidence level and chi-square distribution, which leads to the performance improvement
of the state estimator proposed in Section \ref{secmro} in contrast to the state-of-the-art state estimator proposed in \cite{et1}.
However, the event-triggered schemes provided in \cite{et1} and \cite{et7} do not establish any connection of these parameters.
Also, the proposed event-triggered scheme is novel and different those presented in \cite{et2}$-$\cite{et6} and \cite{et8}$-$\cite{etst3}.
\label{Remarkdev1}
\end{remark}
\par
Let $\hat{x}_k\triangleq \textrm{E}[x_k|\textrm{I}_{k}]$\vspace{3pt} which is the optimal MMSE estimate of $x_k$ given $\textrm{I}_k$.
The rest of the study has two main objectives. The
first objective is to design a MMSE state estimator using the above event-triggered scheme based on confidence level, which can recursively compute $\hat{x}_k$  under a regular assumption.
The second objective is to develop two algorithms for estimating the communication rate of the proposed MMSE state estimator.
\section{ MMSE State Estimation \label{secmro}}
In this section,
we study the MMSE state estimation problem based on the event-triggered scheme. More precisely, the computational strategy for the MMSE estimate $\hat{x}_k$ is studied in the section.
For notational simplicity,
let\vspace{-0.1pt}
\begin{align}
 P_k\triangleq&\textrm{Var}(x_k|\textrm{I}_{k}),\label{notb1}\\
\Omega_k\triangleq &\{z_k\in\mathbb{R}^p|\varphi_k\leq \chi_\alpha^2(p)\}\label{Om},\\
\hat{x}_k^{[z]}\triangleq& \textrm{E}[x_k|\textrm{I}_{k-1},z_k],\label{notc}\\
  P_k^{[z]}\triangleq&\textrm{Var}(x_k|\textrm{I}_{k-1},z_k),\label{notc1}\\
\hat{x}_{k,k-1}\triangleq&\textrm{E}[x_k|\textrm{I}_{k-1}]\label{LeGhna},\\
 \tilde{x}_k\triangleq& x_k-\hat{x}_{k,k-1},\label{notb2}\\
M_k\triangleq& \textrm{Var}(x_k|\textrm{I}_{k-1})\label{LeGhnb},\\
N_k^{[z]}\triangleq &\textrm{Var}(z_k|\textrm{I}_{k-1}),\label{og0k}\\
K_k\triangleq &\textrm{E}\Big[\tilde{x}_k(z_k-\textrm{E}[z_k|\textrm{I}_{k-1}])^\textrm{T}\Big]\big(N_k^{[z]}\big)^{-1}.\label{og0j}
\end{align}
\par
\begin{lemma}
Under the assumption that $f(x_{k-1}|\textrm{I}_{k-1})$ is Gaussian, it holds that:\\
1). $f(z_{k}|\textrm{I}_{k-1})$ is Gaussian.\vspace{1pt}\\
2).  $f(z_{k}|\textrm{I}_{k-1})=\frac{g(z_k)}{(2\pi)^{0.5p}\big|N_k^{[z]}\big|^{0.5}}$ with
\begin{align}
g(z_k)\triangleq \textrm{exp}\big\{\!\!\!-0.5z_k^\textrm{T}(N_k^{[z]})^{-1}z_k\big\}\nonumber.\end{align}
\label{lemfGha}
\end{lemma}
\par
\textit{Proof:} See Appendix \ref{appenT1}.
\begin{lemma}
Under the assumption that $f(x_{k-1}|\textrm{I}_{k-1})$ is Gaussian, it holds\vspace{2pt} that:\\
1). $\int_{\Omega_k}f(z_k|\textrm{I}_{k-1})dz_k=\frac{h_k}{(2\pi)^{0.5p}\big|N_k^{[z]}\big|^{0.5}}$ where
\begin{align}h_k=\int\limits_{-\check{b}}^{\check{b}} dz_{k,1}\!\!\!\!
\int\limits_{-\check{z}_{k,1}}^{\check{z}_{k,1}}\!\!\!\! dz_{k,2}\!\!\!\!
\int\limits_{-\check{z}_{k,2}}^{\check{z}_{k,2}}\!\!\!\! dz_{k,3}\cdots\!\!\!\!\!\! \int\limits_{-\check{z}_{k,p-1}}^{\check{z}_{k,p-1}}\!\!\!\! g(z_k)dz_{k,p}\nonumber\end{align}
with \begin{align}&\check{b}\triangleq \sqrt{\chi_\alpha^2(p)},\check{z}_{k,1}\triangleq \sqrt{\chi_\alpha^2(p)-z_{k,1}^2}, \nonumber\\
&\check{z}_{k,2}\triangleq \sqrt{\chi_\alpha^2(p)-\sum_{i=1}^{2}z_{k,i}^2}, \cdots, \check{z}_{k,p-1}\triangleq \sqrt{\chi_\alpha^2(p)-\sum_{i=1}^{p-1}z_{k,i}^2}\nonumber.\end{align}
2). $\int_{\Omega_k}f(z_k|\textrm{I}_{k-1})z_kz_k^\textrm{T} dz_k=\frac{\varPsi_{k}}{(2\pi)^{0.5p}\big|N_k^{[z]}\big|^{0.5}}$\vspace{3pt}
with $\varPsi_{k}=(\psi_{k,ij})_{p\times p}$ where  $\psi_{k,ij}=\ \! \int\limits_{-\check{b}}^{\check{b}}\!\! dz_{k,1}\!\!\!\!
\int\limits_{-\check{z}_{k,1}}^{\check{z}_{k,1}}\!\!\!\! dz_{k,2}\cdots \!\!\!\!\!\! \int\limits_{-\check{z}_{k,p-1}}^{\check{z}_{k,p-1}}\!\!\!\! g(z_k)z_{k,i}z_{k,j}dz_{k,p}$.
\label{lemfOm}
\end{lemma}\vspace{2pt}
\par
\textit{Proof:} See Appendix \ref{appenT1}.
\par
Considering that $\gamma_k$ has two possible values 0 and 1, we deal with the problem under the following two cases:
\par
1) $\gamma_k=1$. Noticing the definition of $\textrm{I}_{k}$, we have $\textrm{I}_{k}=(\textrm{I}_{k-1},y_k)$.
Then, 
using Kalman filter, we derive
\begin{align}
\hat{x}_k=&\hat{x}_{k,k-1}+M_kC^\textrm{T}(CM_kC^\textrm{T}+R)^{-1}\tilde{y}_k,\label{og1c}\\
P_k=&M_k-M_kC^\textrm{T}(CM_kC^\textrm{T}+R)^{-1}CM_k\label{og1ca}
\end{align}
where $\tilde{y}_k=y_k-C\hat{x}_{k,k-1}$,
$\hat{x}_{k,k-1}=A\hat{x}_{k-1}$ and $M_k=AP_{k-1}A^\textrm{T}+Q$.
\par
2) $\gamma_k=0$. We have $\textrm{I}_{k}=(\textrm{I}_{k-1},\gamma_k=0)$ when $\gamma_k=0$. Then, we present the following theorem to compute $\hat{x}_k$ in $\gamma_k=0$.
\begin{theorem}
\label{Theorem1} When $f(x_{k-1}|\textrm{I}_{k-1})$ is Gaussian, the MMSE state estimation $\hat{x}_k$ and the corresponding error covariance $P_k$ in $\gamma_k=0$ can be computed according to the following equalities:\vspace{-0.1pt}
\begin{align}
\hat{x}_k=&\hat{x}_{k,k-1},\label{The1a}\\
K_k=&M_k C^\textrm{T}(CM_kC^\textrm{T}+R)^{-1}\varPhi^{-1},\label{The1d}\\
P_k^{[z]}=&M_k-M_kC^\textrm{T}(CM_kC^\textrm{T}+R)^{-1}CM_k,\label{The1c}\\
P_k=&P_k^{[z]}+\frac{1}{h_k}K_k\varPsi_{k} K_k^\textrm{T}.\label{The1P}
\end{align}
\end{theorem}
\par \emph{Proof:} See Appendix \ref{appenT3}.
\begin{remark}
In fact, $\hat{x}_k$ in (\ref{The1a}) of Theorem \ref{Theorem1} \vspace{3pt}should be computed via $\hat{x}_k=\hat{x}_{k,k-1}+e_k$\vspace{3pt}
with $e_k\triangleq \frac{K_k\int_{\Omega_k} f(z_k|\textrm{I}_{k-1}) z_k dz_k }{\int_{\Omega_k}f(z_k|\textrm{I}_{k-1})dz_k}$ in which
 $\int_{\Omega_k} f(z_k|\textrm{I}_{k-1}) z_k dz_k$ can be obtained via \vspace{3pt} $\int_{\Omega_k} f(z_k|\textrm{I}_{k-1}) z_k dz_k=\frac{\psi_{k}}{(2\pi)^{0.5p}\big|N_k^{[z]}\big|^{0.5}}$
where\vspace{3pt} $\psi_{k}=(\psi_{k,1},\psi_{k,2},\cdots,\psi_{k,p})^\textrm{T}$ with $\psi_{k,i}=\ \! \int\limits_{-\check{b}}^{\check{b}}\!\!\!\! dz_{k,1}\!\!\!\!\vspace{3pt}
\int\limits_{-\check{z}_{k,1}}^{\check{z}_{k,1}}\!\!\!\! dz_{k,2}\cdots \!\!\!\!\!\! \int\limits_{-\check{z}_{k,p-1}}^{\check{z}_{k,p-1}}\!\!\!\! g(z_k)z_{k,i}dz_{k,p}$, $i=1,2,\cdots,p$. Since $\psi_{k}$ is almost equal to zero vector, we conclude that
$e_k$ is almost equal to zero vector. Hence, $\hat{x}_k$ in  Theorem \ref{Theorem1} is calculated using (\ref{The1a}).
\label{Remarkexk}
\end{remark}
\par
Now, we can present the MMSE state estimator.
\\
Starting with $\hat{x}_{k-1}$ and $P_{k-1}$, the MMSE state estimator includes the following two steps.
\\
\textbf{Step 1:} Compute $\hat{x}_{k,k-1}$, $\tilde{y}_k$, $M_k$, $h_k$ and $\varPsi_{k}$ according to
\begin{align}
\hat{x}_{k,k-1}=&A \hat{x}_{k-1},\label{Al1a0}\\
\tilde{y}_k=&y_k-C\hat{x}_{k,k-1},\label{Al1aa0}\\
M_k=&AP_{k-1}A^\textrm{T}+Q,
\label{Al1a}\\
N_k^{[z]}=&\varPhi(CM_kC^\textrm{T}+R)\varPhi^\textrm{T},\label{Al1b}\\
h_k=&\int\limits_{-\check{b}}^{\check{b}}\!\!\! dz_{k,1}\!\!\!\!
\int\limits_{-\check{z}_{k,1}}^{\check{z}_{k,1}}\!\!\!\! dz_{k,2}\cdots\!\!\!\!\!\! \int\limits_{-\check{z}_{k,p-1}}^{\check{z}_{k,p-1}}\!\!\!\! g(z_k)dz_{k,p},\label{St1a}\\
\psi_{k,ij}=&\ \! \int\limits_{-\check{b}}^{\check{b}}\!\!\! dz_{k,1}\!\!\!\!
\int\limits_{-\check{z}_{k,1}}^{\check{z}_{k,1}}\!\!\!\! dz_{k,2}\cdots\!\!\!\!\!\! \int\limits_{-\check{z}_{k,p-1}}^{\check{z}_{k,p-1}}\!\!\!\! g(z_k)z_{k,i}z_{k,j}dz_{k,p},\label{St1b}\\
\varPsi_{k}=&(\psi_{k,ij})_{p\times p}\label{St1c}
\end{align}
where $g(z_k)$ is defined in 2) of Lemma \ref{lemfGha},  as well as $\check{b}$ and $\check{z}_{k,i}$ with $i=1,2,\cdots,p-1$ are defined in 1) of Lemma \ref{lemfOm}.
\\
\textbf{Step 2:} Compute $\hat{x}_k$ and $P_k$ in terms of
\begin{align}
P_k^{[z]}=&M_k-M_kC^\textrm{T}(CM_kC^\textrm{T}+R)^{-1}CM_k,\label{Al1c}\\
\hat{x}_k=&\hat{x}_{k,k-1}+\gamma_kM_kC^\textrm{T}(CM_kC^\textrm{T}+R)^{-1}\tilde{y}_k,\label{Al1d}\\
K_k=&M_k C^\textrm{T}(CM_kC^\textrm{T}+R)^{-1}\varPhi^{-1},
\label{Al1e}\\
P_k=&P_k^{[z]}+\frac{(1-\gamma_k)}{h_k}K_k\varPsi_{k} K_k^\textrm{T}\label{Al1f}
\end{align}
where $\gamma_k$ is determined by (\ref{et}) and (\ref{ySy}).
For the proposed MMSE state estimator, we easily see that we only need to prove (\ref{Al1d}) and (\ref{Al1f}) where we easily obtain  (\ref{Al1d}) by using (\ref{og1c}) and (\ref{The1a}). Using
(\ref{og1ca}) and (\ref{The1c}), we see that\vspace{-3.5pt}
\begin{align}
P_k=P_k^{[z]} \ \ \ \textrm{when} \ \gamma_k=1.\label{Al1fa}
\end{align}
Putting  (\ref{Al1fa}) and (\ref{The1P}) together, we prove (\ref{Al1f}).
\begin{remark}
The results for MMSE state estimation problem based on different event-triggered schemes were presented in \cite{et1}, \cite{et1a}, \cite{et5}, \cite{et6}, \cite{et8}, \cite{et9},
\cite{et11}, \cite{et12}, \cite{etad1} and \cite{etst3}.
However,  compared with the results,
the MMSE state estimator presented in this paper has a different strategy in computing the error covariance $P_k$
because a novel confidence level based event-triggered scheme is applied to the design of the MMSE state estimator.
\label{Remarko1}
\end{remark}
\par
\begin{remark}
For the MMSE state estimator presented in (\ref{Al1a0})$-$(\ref{Al1f}), we need to know the initial conditions $\hat{x}_{0}$ and $P_{0}$.
Hence, we present the following scheme to obtain $\hat{x}_{0}$ and $P_{0}$.
\label{Remarkin}
\end{remark}
\par
$\hat{x}_{0}$ and $P_{0}$ can be computed according to
\begin{align}
\hat{x}_0=&\bar{x}_{0}+\gamma_0\bar{P}_0C^\textrm{T}(C\bar{P}_0C^\textrm{T}+R)^{-1}(y_0-C\bar{x}_{0}),\label{0ina}\\
\bar{N}_0^{[z]}=&\varPhi(C\bar{P}_0C^\textrm{T}+R)\varPhi^\textrm{T},\label{0inc}
\end{align}
\begin{align}
\bar{\psi}_{0,ij}=&\ \! \int\limits_{-\check{b}}^{\check{b}}\!\!\! dz_{0,1}\!\!\!\!
\int\limits_{-\check{z}_{0,1}}^{\check{z}_{0,1}}\!\!\!\! dz_{0,2}\cdots\!\!\!\!\!\! \int\limits_{-\check{z}_{0,p-1}}^{\check{z}_{0,p-1}}\!\!\!\!
\rho(z_0)z_{0,i}z_{0,j}dz_{0,p},\label{0inb}\\
\bar{\varPsi}_0=&(\bar{\psi}_{0,ij})_{p\times p},\label{0inb1}\\
\alpha_0=&\int\limits_{-\check{b}}^{\check{b}}\!\!\! dz_{0,1}\!\!\!\!
\int\limits_{-\check{z}_{0,1}}^{\check{z}_{0,1}}\!\!\!\! dz_{0,2}\cdots\!\!\!\!\!\! \int\limits_{-\check{z}_{0,p-1}}^{\check{z}_{0,p-1}}\!\!\!\!
\rho(z_0)dz_{0,p},\label{0ind}\\
\bar{K}_0=&\bar{P}_0C^\textrm{T}(C\bar{P}_0C^\textrm{T}+R)^{-1}\varPhi^{-1},\label{0ine}\\
P_0^{[z]}=&\bar{P}_0-\bar{P}_0C^\textrm{T}(C\bar{P}_0C^\textrm{T}+R)^{-1}C\bar{P}_0,\label{0inf}\\
P_0=&{P}_0^{[z]}+\frac{(1-\gamma_0)}{\alpha_0}\bar{K}_0 \bar{\varPsi}_0\bar{K}_0^\textrm{T}\label{0ing}
\end{align}
where $\gamma_0$ is determined by (\ref{et}) with $\varphi_0=(y_0-C\bar{x}_0)^\textrm{T}\varSigma(y_0-C\bar{x}_0)$, as well as
 $\alpha_0\triangleq\int_{\Omega_{0}}\rho(z_0)dz_{0}$, $\rho(z_0)\triangleq\textrm{exp}\big\{\!\!\!-0.5z_0^\textrm{T}(\bar{N}_0^{[z]})^{-1}z_0\big\}$, $\bar{N}_0^{[z]}\triangleq \textrm{Var}(z_0)$ and $
\bar{K}_0\triangleq\textrm{Cov}(x_0,z_0)\big(\bar{N}_0^{[z]}\big)^{-1}$.
Making reference to the derivation of the MMSE state estimator, we can easily obtain (\ref{0ina})$-$(\ref{0ing}).
\section{Communication Rate Estimation \label{sbd}}
In this section, we study the communication rate estimation problem for the proposed  MMSE state estimator. More precisely, we will present two strategies for approximately computing
$\textrm{E}[\gamma_k]$.
\par
$\textrm{E}[\gamma_k]$ can be expressed as\vspace{-0.1pt}
\begin{align}
\textrm{E}[\gamma_k]=&0\times P(\gamma_k=0)+1\times P(\gamma_k=1)\nonumber\\
=&P(\gamma_k=1)=1-P(\gamma_k=0)\label{Ega0}
\end{align}
where
\begin{align}
P(\gamma_k=0)=&\int f(\gamma_k=0,\textrm{I}_{k-1})d\textrm{I}_{k-1}
\nonumber\\
=&\int f(\textrm{I}_{k-1}) P(\gamma_k=0|\textrm{I}_{k-1})d\textrm{I}_{k-1}.
\label{Ega2a1}
\end{align}
\begin{remark}
From (\ref{Ega2a1}), we see that the computation of $P(\gamma_k=0)$ is intractable because the computational complex of $\int f(\textrm{I}_{k-1}) P(\gamma_k=0|\textrm{I}_{k-1})d\textrm{I}_{k-1}$ increases with $k$. Then, it follows from
(\ref{Ega0}) that the computation of $\textrm{E}[\gamma_k]$ is intractable. In order to approximately compute $\textrm{E}[\gamma_k]$,
we will use two types of approximations where one type of approximation is $\textrm{E}[\gamma_k]\approx \textrm{E}[\gamma_k|\textrm{I}_{k-1}]$, and the other one is
$\textrm{E}[\gamma_k]\approx \textrm{E}[\gamma_k|\textrm{I}_{k-2}]$. We will present a strategy for computing $\textrm{E}[\gamma_k|\textrm{I}_{k-1}]$ and $\textrm{E}[\gamma_k|\textrm{I}_{k-2}]$, and
we will test the two different approximations for $\textrm{E}[\gamma_k]$ in Section  \ref{sectionNEa}.
\label{Remarkcr}
\end{remark}
\par
For notational simplicity, let
\begin{align}
\hat{\gamma}_{k,k-i}\triangleq& \textrm{E}[\gamma_k|\textrm{I}_{k-i}],\label{Noe0a}\\
P_{k,k-i}(0)\triangleq& P(\gamma_k=0|\textrm{I}_{k-i}),\label{Nop0}\\
\vec{P}_k^{[z]}(0)\triangleq&P(\gamma_k=0|\textrm{I}_{k-2},z_{k-1}),\label{Nop1a}\\
\breve{P}_k(0)\triangleq&P(\gamma_k=0|\textrm{I}_{k-2},\gamma_{k-1}=0),\label{Nop1aA}\\
\hat{z}_{k,k-1}^\triangleright\triangleq &\textrm{E}[z_{k}|\textrm{I}_{k-2},z_{k-1}],\label{Nca0}\\
\tilde{z}_{k}^{\triangleright}\triangleq &z_{k}-\hat{z}_{k,k-1}^\triangleright,\label{Nca5}\\
\vec{N}_k^{[z]}\triangleq &\textrm{Var}(z_k|\textrm{I}_{k-2},z_{k-1}),\label{Nca1}\\
\hat{z}_{k,k-1}(0)\triangleq &\textrm{E}[z_{k}|\textrm{I}_{k-2},\gamma_{k-1}=0],\label{Nca6}\\
N_k(0)\triangleq &\textrm{Var}(z_k|\textrm{I}_{k-2},\gamma_{k-1}=0),\label{Nca7}
\end{align}
\begin{align}
\breve{\Omega}_k\triangleq &\{z_k\in\mathbb{R}^p|\varphi_k> \chi_\alpha^2(p)\}\label{Omc}
\end{align}
with $i=1,2$.
\par
Starting with $\hat{x}_{k-1}$ and $P_{k-1}$, $\hat{\gamma}_{k,k-1}$ can be recursively computed according to Algorithm \ref{Pa1}.
\begin{algorithm}[htb]
\setstretch{1.1} 
\caption{\textbf{: Communication rate based on information up to $k-1$}}
\label{Pa1}
\textbf{Step 1:} Compute $\hat{\gamma}_{k,k-1}$  according to
\begin{align}
P_{k,k-1}(0)=&\frac{h_k}{(2\pi)^{0.5p}|N_k^{[z]}|^{0.5}},\label{Alo4a}\\
\hat{\gamma}_{k,k-1}=&1-P_{k,k-1}(0)\label{Alo4b}
\end{align}
where $N_k^{[z]}$ and $h_k$ are computed using (\ref{Al1a})$-$(\ref{St1a}) in sequence.
\\
\textbf{Step 2:}  Compute and store $\hat{x}_{k}$ and $P_{k}$ for the derivation of $\hat{\gamma}_{k+1,k}$ where
 $\hat{x}_{k}$ and $P_{k}$ are computed via  (\ref{Al1a0})$-$(\ref{Al1f}) in sequence.
%
%
%
%
%
\end{algorithm}
For Algorithm  \ref{Pa1}, we only need to prove (\ref{Alo4a}) and (\ref{Alo4b}).
$P(\gamma_k=0|\textrm{I}_{k-1})$ can be rewritten as
 \begin{align}
P(\gamma_k=0|\textrm{I}_{k-1})=&P(z_k\in\Omega_{k}|\textrm{I}_{k-1})\nonumber\\
=&\int_{\Omega_{k}}f(z_{k}|\textrm{I}_{k-1})dz_{k}
=\frac{h_k}{(2\pi)^{0.5p}\big|N_k^{[z]}\big|^{0.5}}\label{Pa1a}
\end{align}
where $\Omega_{k}$ is defined in  (\ref{Om}), and the last equality is due to 1) of Lemma \ref{lemfOm}.
Then, replacing $P(\gamma_k=0|\textrm{I}_{k-1})$ by $P_{k,k-1}(0)$, we prove (\ref{Alo4a}). Making reference to (\ref{Ega0}), we easily obtain (\ref{Alo4b}).
\par
In order to compute $\hat{\gamma}_{k,k-2}$, we propose the following theorem.
\begin{theorem}
\label{Theorem2} When $f(x_{k-1}|\textrm{I}_{k-1})$ is Gaussian, $\hat{\gamma}_{k,k-2}$ can be computed in terms of the following equalities:
\begin{align}
\vec{N}_k^{[z]}
=&\varPhi\big(C(AP_{k-1}^{[z]} A^\textrm{T}+Q) C^\textrm{T}+R)\varPhi^\textrm{T},\label{The2b}\\
N_k(0)=\varPhi&\Big(C\big(A(P_{k-1}^{[z]}+\frac{1}{h_{k-1}}K_{k-1}\varPsi_{k-1} K_{k-1}^\textrm{T})A^\textrm{T}+Q\big)C^\textrm{T}+R\Big)\nonumber\\
&\times\varPhi^\textrm{T},\label{The2b1}\\
f(z_{k}|\textrm{I}&_{k-2},z_{k-1})=\frac{\vec{g}(z_k)}{(2\pi)^{0.5p}\big|\vec{N}_k^{[z]}\big|^{0.5}},\label{The2c}\\
f(z_{k}|\textrm{I}&_{k-2},z_{k-1})=\frac{\vec{g}(z_k)}{(2\pi)^{0.5p}\big|\vec{N}_k^{[z]}\big|^{0.5}},\label{The2c}\\
f(z_{k}|\textrm{I}&_{k-2},\gamma_{k-1}=0)=\frac{\breve{g}(z_k)}{(2\pi)^{0.5p}\big|N_k(0)\big|^{0.5}},\label{The2ca}\\
\vec{P}_k^{[z]}(0)=&\int_{\Omega_{k}}f(z_{k}|\textrm{I}_{k-2},z_{k-1})dz_{k},\label{The2d}\\
\breve{P}_k(0)=&\int_{\Omega_{k}}f(z_{k}|\textrm{I}_{k-2},\gamma_{k-1}=0)dz_{k},\label{The2da}\\
{P}_{k,k-2}(0)=&P_{k-1,k-2}(0)\breve{P}_k(0)+
 \int_{\breve{\Omega}_{k-1}}\vec{P}_k^{[z]}(0)f(z_{k-1}|\textrm{I}_{k-2})dz_{k-1},\label{The2e}\\
 \hat{\gamma}_{k,k-2}=&1-{P}_{k,k-2}(0)\label{The2f}
\end{align}
where\vspace{-0.1pt}
\begin{align}
\vec{g}(z_k)\triangleq\textrm{exp}\big\{\!\!\!-0.5{z}_k^\textrm{T}(\vec{N}_k^{[z]})^{-1}{z}_k\big\},\nonumber\\
\breve{g}(z_k)\triangleq\textrm{exp}\big\{\!\!\!-0.5{z}_k^\textrm{T}N_k(0)^{-1}{z}_k\big\}.\label{Nca4}
\end{align}
\end{theorem}
\par
\emph{Proof:} See Appendix \ref{appenT4}.
\par
Based on the above discussion, we present an algorithm to compute $\hat{\gamma}_{k,k-2}$ in a recursive structure.
Starting with $P_{k-1}^{[z]}$, $h_{k-1}$, $K_{k-1}$, $\varPsi_{k-1}$, $\hat{x}_{k-1}$, $P_{k-1}$ and $P_{k-1,k-2}(0)$, $\hat{\gamma}_{k,k-2}$ can be recursively computed according to Algorithm \ref{Pa2} where
the proof of Algorithm \ref{Pa2} is presented in Appendix \ref{appenT5}.
\begin{algorithm}[htb]
\setstretch{1.1} 
\caption{\textbf{: Communication rate based on information up to $k-2$ }}
\label{Pa2}
\textbf{Step 1:} Compute  $\vec{N}_k^{[z]}$ and $N_k(0)$ using (\ref{The2b}) and (\ref{The2b1}), respectively.\vspace{2pt}
 \\
\textbf{Step 2:}  Compute $\vec{P}_k^{[{z}]}(0)$ and $\breve{P}_k(0)$ according to
\begin{align}
\vec{P}_k^{[{z}]}(0)=& \frac{\int\limits_{-\check{b}}^{\check{b}}\!\!\! dz_{k,1}\!\!\!\!
\int\limits_{-\check{z}_{k,1}}^{\check{z}_{k,1}}\!\!\!\! dz_{k,2}
\cdots\!\!\!\!\!\! \int\limits_{-\check{z}_{k,p-1}}^{\check{z}_{k,p-1}}\!\!\!\! \vec{g}(z_k)dz_{k,p}}
{(2\pi)^{0.5p}\big|\vec{N}_k^{[z]}\big|^{0.5}},
 \label{Al2d}\\
 \breve{P}_k(0)=& \frac{\int\limits_{-\check{b}}^{\check{b}}\!\!\! dz_{k,1}\!\!\!\!
\int\limits_{-\check{z}_{k,1}}^{\check{z}_{k,1}}\!\!\!\! dz_{k,2}
\cdots\!\!\!\!\!\! \int\limits_{-\check{z}_{k,p-1}}^{\check{z}_{k,p-1}}\!\!\!\! \breve{g}(z_k)dz_{k,p}}
{(2\pi)^{0.5p}|N_k(0)|^{0.5}}\label{Al2e}
\end{align}
where $\vec{g}(z_k)$ and $\breve{g}(z_k)$ are defined in (\ref{Nca4}).
\\
\textbf{Step 3:} Compute ${P}_{k,k-2}(0)$ using
\begin{align}
 {P}_{k,k-2}(0)
 =&\vec{P}_k^{[z]}(0)+P_{k-1,k-2}(0)\big(\breve{P}_k(0)-\vec{P}_k^{[z]}(0)\big).\label{Al2f}
\end{align}
\textbf{Step 4:} Compute $\hat{\gamma}_{k,k-2}$ in terms of (\ref{The2f}).
\\
\textbf{Step 5:} For computing the communication rate at time step $k+1$, update  $P_{k}^{[z]}$, $h_{k}$, $K_{k}$, $\varPsi_{k}$, $\hat{x}_{k}$, $P_{k}$ and $P_{k,k-1}(0)$ using the MMSE state estimator presented in (\ref{Al1a0})$-$(\ref{Al1f}) and using (\ref{Alo4a}).
%
%
%
%
%
\end{algorithm}
\begin{remark}
Under the assumption that $f(x_{k-1}|\textrm{I}_{k-1})$ is Gaussian, we obtain
$\hat{\gamma}_{k,k-2}$ using Algorithm \ref{Pa2} in a recursive form where the proof of the algorithm is very challenging.
In order to prove Algorithm \ref{Pa2},  Lemma \ref{lemyeqz} and Theorem \ref{Theorem2} are first proved in Appendix \ref{appenT4}, and then Algorithm \ref{Pa2}
 is proved in Appendix \ref{appenT5}.
\label{Remarkada2a}
\end{remark}
\begin{remark}
If we take the approximation $\textrm{E}[\gamma_k]\approx \hat{\gamma}_{k,k-1}$ where $\hat{\gamma}_{k,k-1}$ can be obtained from
Algorithm \ref{Pa1}, we need to additionally obtain $\textrm{E}[\gamma_0]$ since Algorithm \ref{Pa1} starts with $k=1$.
In the same way, we need to know $\textrm{E}[\gamma_0]$ and  $\textrm{E}[\gamma_1]$ if we take the approximation $\textrm{E}[\gamma_k]\approx \hat{\gamma}_{k,k-2}$
using Algorithm \ref{Pa2}. Hence, we need to obtain $\textrm{E}[\gamma_0]$ and $\textrm{E}[\gamma_1]$.
\label{Remarkine}
\end{remark}
We present a strategy for computing $\textrm{E}[\gamma_0]$ and $\textrm{E}[\gamma_1]$  with the following content:
\begin{align}
P(\gamma_0=0)=&\frac{\alpha_0}{(2\pi)^{0.5p}|\bar{N}_0^{[z]}|^{0.5}},\label{0inh}\\
\textrm{E}[\gamma_0]=&1-P(\gamma_0=0),\label{0inha}\\
\bar{N}_1^{[z]}
=&\varPhi\big(C(AP_{0}^{[z]} A^\textrm{T}+Q) C^\textrm{T}+R)\varPhi^\textrm{T},\label{0ini}\\
\bar{N}_1(0)=&\varPhi\Big(C\big(A({P}_0^{[z]}+\frac{1}{\alpha_0}\bar{K}_0\bar{\varPsi}_0 \bar{K}_0^\textrm{T})A^\textrm{T}+Q\big)C^\textrm{T}+R\Big)\varPhi^\textrm{T},\label{0inj}
\end{align}
\begin{align}
\bar{P}_1(0)=&\frac{\int\limits_{-\check{b}}^{\check{b}}\!\!\! dz_{1,1}\!\!\!\!
\int\limits_{-\check{z}_{1,1}}^{\check{z}_{1,1}}\!\!\!\! dz_{1,2}
\cdots\!\!\!\!\!\! \int\limits_{-\check{z}_{1,p-1}}^{\check{z}_{1,p-1}}\!\!\!\! \breve{\rho}(z_1)dz_{1,p}}
{(2\pi)^{0.5p}\big|\bar{N}_1(0)\big|^{0.5}},\label{0ink}\\
\bar{P}_1^{[z]}(0)=&\frac{\int\limits_{-\check{b}}^{\check{b}}\!\!\! dz_{1,1}\!\!\!\!
\int\limits_{-\check{z}_{1,1}}^{\check{z}_{1,1}}\!\!\!\! dz_{1,2}
\cdots\!\!\!\!\!\! \int\limits_{-\check{z}_{1,p-1}}^{\check{z}_{1,p-1}}\!\!\!\! \vec{\rho}(z_1)dz_{1,p}}
{(2\pi)^{0.5p}\big|\bar{N}_1^{[z]}\big|^{0.5}},\label{0inl}\\
P(\gamma_1=0)=&\bar{P}_1^{[z]}(0)+P(\gamma_0=0)(\bar{P}_1(0)-\bar{P}_1^{[z]}(0)),\label{0inm}\\
\textrm{E}[\gamma_1]=&1-P(\gamma_1=0)\label{0inn}
\end{align}
where $\bar{N}_0^{[z]}$, $\alpha_0$ and $P_{0}^{[z]}$ are computed using \vspace{2pt} (\ref{0inc}), (\ref{0ind}) and  (\ref{0inf}); and\vspace{2pt} $\bar{N}_1(0)\triangleq \textrm{Var}(z_1|\gamma_0=0)
$, $\bar{N}_1^{[z]}\triangleq\textrm{Var}(z_1|z_0)$, $\breve{\rho}(z_1)\triangleq\textrm{exp}\big\{\!\!\!-0.5z_1^\textrm{T}\big(\bar{N}_1(0)\big)^{-1}z_1\big\}$,
$\vec{\rho}(z_1)\triangleq\textrm{exp}\big\{\!\!\!-0.5z_1^\textrm{T}\big(\bar{N}_1^{[z]}\big)^{-1}z_1\big\}$,\vspace{2pt} $\bar{P}_1(0)\triangleq P(\gamma_1=0|\gamma_{0}=0)$
and $\bar{P}_1^{[z]}(0)\triangleq P(\gamma_1=0|z_{0})$.\vspace{1pt}
Making reference to the derivation of Algorithm \ref{Pa1}, we easily derive (\ref{0inh}). Using (\ref{Ega0}), we directly derive
(\ref{0inha}) and (\ref{0inn}).  Similar to the derivation of Algorithm \ref{Pa2}, we easily
obtain (\ref{0ini})$-$(\ref{0inm}).
\section{Simulation Example \label{sectionNE}}
In this section, we illustrate the performance of the MMSE state estimator and the communication rate estimation algorithms proposed in this paper via a target tracking scenario including two parts.
More precisely, we test the performance of the proposed results in Section \ref{sectionNEa}, and we compare the MMSE state estimator proposed in this paper with the state estimator proposed in
\cite{et1} in Section \ref{sectionNEb}. The state estimator proposed in
\cite{et1} is referred as SEHI considering that its event-triggered scheduler is based on H\"{o}lder infinity-norm.
\subsection{Performance Evaluation of the Proposed Results\label{sectionNEa}}
Consider a target tracking problem \cite{siex} where the state-space
formulation of the target can be written as  (\ref{sxk}) with\vspace{3pt}\\
$x_k=\left(
  \begin{array}{c}
    p_{k} \\
    v_{k} \\
    a_k\\
  \end{array}
\right),$
$A=\left(
          \begin{array}{ccc}
            1 & T & T^2 \\
            0 & 1 & T \\
            0 & 0 & 1 \\
          \end{array}
        \right)$,\vspace{6pt}\\ $Q=2a\sigma_m^2\left(
          \begin{array}{ccc}
            T^5/20 & T^4/8  & T^3/6 \\
            T^4/8 & T^3/3   & T^2/2 \\
            T^3/6 & T^2/2   & T \\
          \end{array}
        \right)$.\vspace{3pt}
\\
$p_{k}$, $v_{k}$ and $a_k$ stand for the position, velocity and acceleration, respectively, of the target. $T$, $a$ and $\sigma_m^2$ denote
the sampling period, the maneuver time constant's reciprocal and the target acceleration's variance, respectively. The measurement of the target can be
modeled as
\vspace{3pt} (\ref{syk})
where
$C=\left(
          \begin{array}{ccc}
            1 & 0 & 0 \\
            0 & 0 & 1 \\
          \end{array}
        \right)$ and $R=\left(
          \begin{array}{cc}
            60 & 0  \\
            0 & 10  \\
          \end{array}
        \right)$. The initial position, velocity and acceleration of this target are 3410m, 30m/s and 0m/s$^2$, respectively. \vspace{3pt}  We select\\
      $  \bar{x}_0=\left(
                                                                                        \begin{array}{c}
                                                                                          3500 \\
                                                                                          40 \\
                                                                                          0\\
                                                                                        \end{array}
                                                                                      \right)
$, $\bar{P}_0=\left(
                   \begin{array}{ccc}
                     60^2 & 60^2/T & 0\\
                     60^2/T & 2\times60^2/T^2 &0 \\
                     0 & 0&  0\\
                   \end{array}
                 \right)
$.\vspace{3pt} \\ Also, we take $T=1$, $a=2$ and $\sigma_m^2=0.5$. We select $1-\alpha=0.95$ in confidence level. Then, noticing $p=2$ and using the chi-square distribution table,
we have $\chi_\alpha^2(p)=5.991$.
\begin{figure}
\vspace{-15pt}
  \begin{center}
\includegraphics[ width=3.4in]{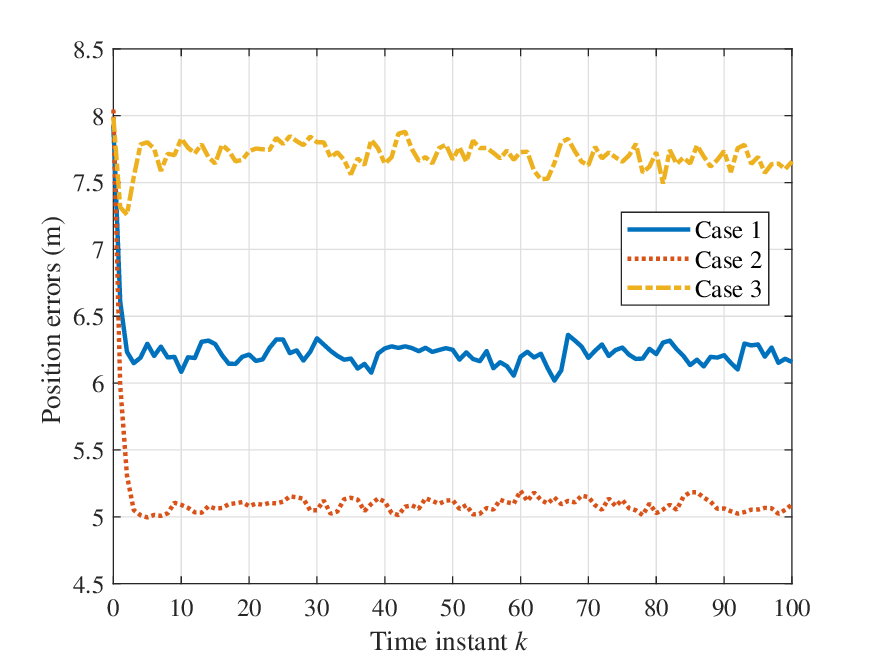}
\caption{  RMS Position errors of SECL for three different cases.}
  \label{figp}
  \end{center}
\end{figure}
\begin{figure}
\vspace{-15pt}
  \begin{center}
\includegraphics[ width=3.4in]{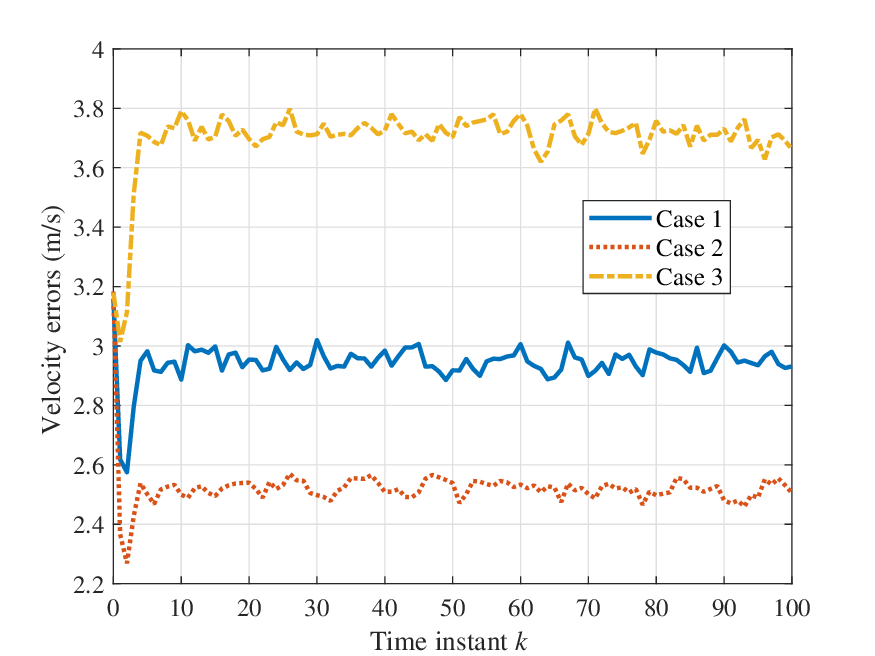}
\caption{    RMS velocity errors of SECL for three different cases.}
  \label{figv}
  \end{center}\vspace{-15pt}
\end{figure}
For the tolerable upper bound $\overline{N}$, we take three different parameter values given by the following three cases:\vspace{2pt}\\
Case 1: $\overline{N}=\left(
                        \begin{array}{cc}
                          50 & 4\\
                          4 & 8 \\
                        \end{array}
                      \right)
$; Case 2: $\overline{N}=0.5\times\left(
                        \begin{array}{cc}
                          50 & 4\\
                          4 & 8 \\
                        \end{array}
                      \right)\vspace{4pt}
$; Case 3: $\overline{N}=\left(
                        \begin{array}{cc}
                          60 & 10\\
                          10 & 20 \\
                        \end{array}
                      \right)
$.\vspace{3pt}
\par
\begin{figure}
\vspace{-15pt}
  \begin{center}
\includegraphics[ width=3.4in]{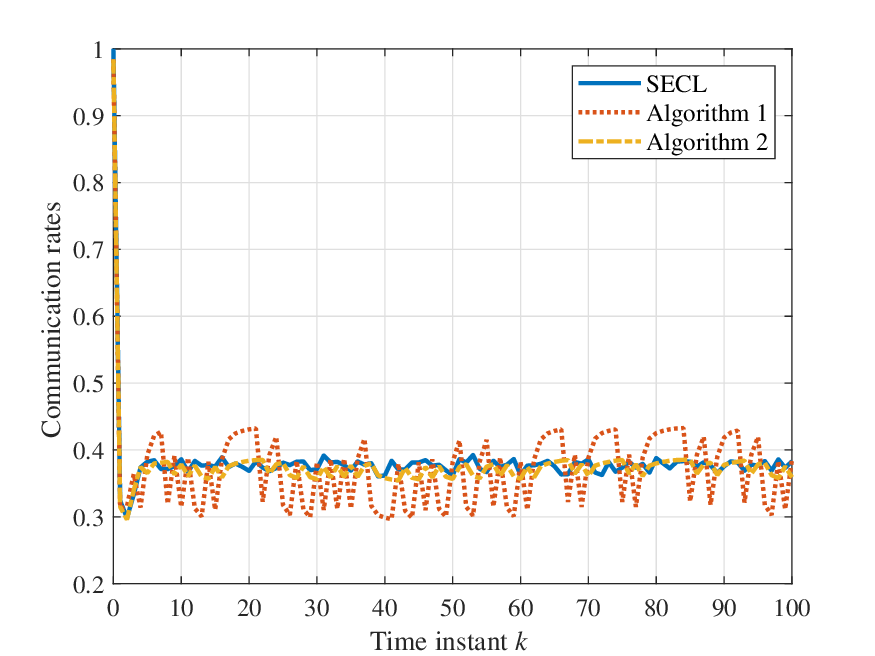}
\caption{  Communication rates of SECL, Algorithm \ref{Pa1} and Algorithm \ref{Pa2} at Case 1.}
  \label{figCRca1}
  \end{center}\vspace{-18pt}
\end{figure}
\begin{figure}
\vspace{-15pt}
  \begin{center}
\includegraphics[width=3.4in]{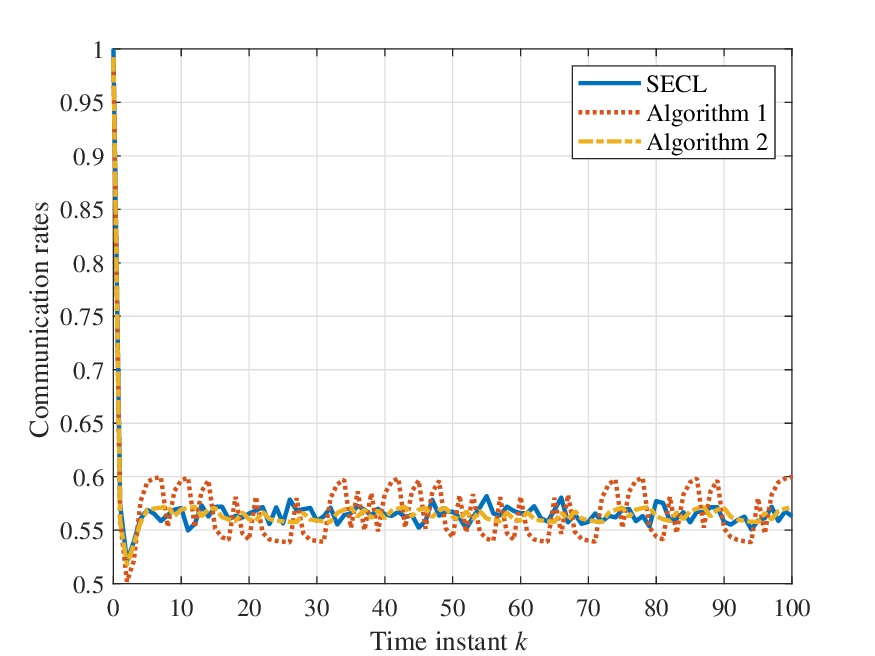}
\caption{  Communication rates of SECL, Algorithm \ref{Pa1} and Algorithm \ref{Pa2} at Case 2.}
  \label{figCRca2}
  \end{center}
\end{figure}
\begin{figure}
\vspace{-15pt}
  \begin{center}
\includegraphics[width=3.4in]{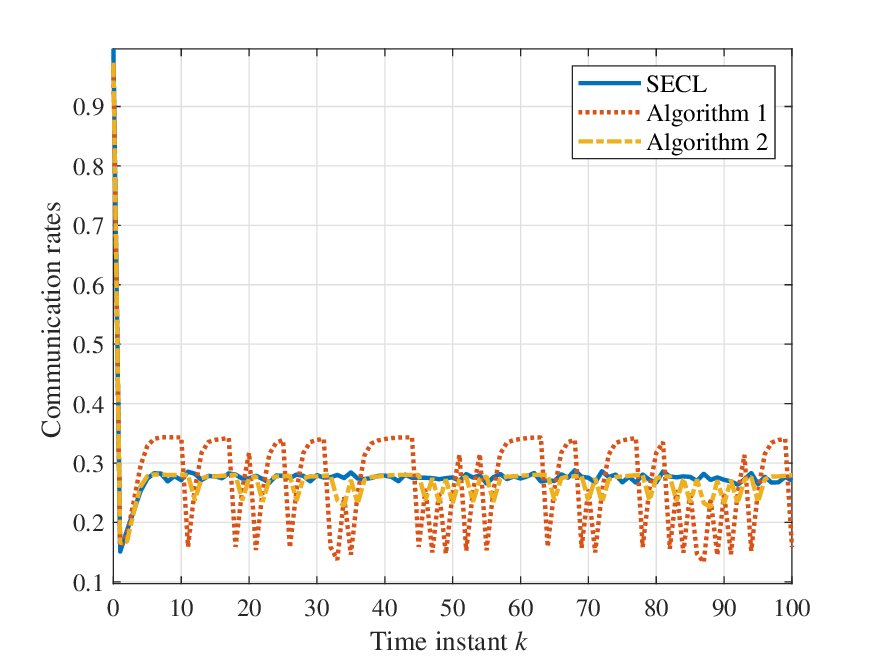}
\caption{  Communication rates of SECL, Algorithm \ref{Pa1} and Algorithm \ref{Pa2} at Case 3.}
  \label{figCRca3}
  \end{center}
\end{figure}

We test the performance of the presented results using a Monte Carlo simulation with $N=5000$ trials, and we take $k=0,1,\cdots,100$ for each trial.
We use the root-mean-square (RMS) error, the communication rate and the average communication rate as the performance evaluation criteria.
At time step $k$, the RMS error is defined as $\sqrt{\frac{1}{N}\sum_{i=1}^N(\varsigma_{k,i}-\hat{\varsigma}_{k,i})^2}$ for $N$ trials where $\varsigma_{k,i}$
stands for the state of $\varsigma_{k}$ at the $i$th trial, and $\hat{\varsigma}_{k,i}$ stands for an estimate of $\varsigma_{k,i}$. Let $\gamma_{k,i}$
denote the state of $\gamma_{k}$ at the $i$th trial, and
the communication rate for the proposed MMSE state estimator at time step $k$ can be computed using the approximation $\textrm{E}[\gamma_{k}]\approx\frac{1}{N}\sum_{i=1}^N\gamma_{k,i}$ where $\textrm{E}[\gamma_{k}]=\frac{1}{N}\sum_{i=1}^N\gamma_{k,i}$
when $N$ approaches infinity.\vspace{2pt} The average communication rate is defined as\vspace{2pt} $\gamma=\lim\limits_{k\rightarrow\infty}\frac{1}{k}\sum_{j=0}^{k-1}\textrm{E}[\gamma_{j}]$, and the average communication rate
is approximately computed via $\gamma\approx\frac{1}{101}\sum_{j=0}^{100}\textrm{E}[\gamma_{j}]$ in the Monte Carlo simulation.
The MMSE state estimator based on confidence level proposed in this paper is referred to as SECL.
The RMS position and velocity errors of SECL for three different cases are given in Figs. \ref{figp} and \ref{figv}, respectively.
The communication rates of SECL, Algorithm \ref{Pa1} and Algorithm \ref{Pa2} at Cases 1, 2 and 3 are provided in Figs. \ref{figCRca1}, \ref{figCRca2} and \ref{figCRca3}, respectively, where, without loss of generality, we take the information at the 40th trial for running
Algorithms \ref{Pa1} and \ref{Pa2}.
From Figs. \ref{figp}$-$\ref{figCRca3}, we find that, for both position and velocity estimate at the three different cases,
SECL has different performances, and
the performance is connected with the communication rate. More precisely, the performance of SECL becomes better and better with the increase of communication rate.
This indicates the effectiveness of the MMSE state estimator based on confidence level proposed in this paper.
Figs. \ref{figCRca1}$-$\ref{figCRca3} also shows that Algorithm \ref{Pa2} provides a good estimate for the communication rate of SECL because the communication rate yielded by
Algorithm \ref{Pa2} is close to the communication rate of SECL. The average communication rates of SECL, Algorithm \ref{Pa1} and Algorithm \ref{Pa2} for the three cases are provided in Table \ref{tab1}.
We see from Table \ref{tab1} that the average communication rates of SECL, Algorithm \ref{Pa1} and Algorithm \ref{Pa2} are very close at all the three cases, which means that both Algorithm  \ref{Pa1} and Algorithm \ref{Pa2}
yield good performance for estimating the average communication rates of SECL.
\par
Hence, the simulation results indicate that SECL is effective in solving state estimation with the trade-off between communication rate and state estimation performance, and that Algorithm \ref{Pa2} yields a good performance in estimating
the communication rate and the average communication rate of SECL.
\begin{table}[!t]
\begin{center}
\caption{Average communication rates of SECL, Algorithm \ref{Pa1} and Algorithm \ref{Pa2}
}
 \label{tab1}
\centering
\vspace{5pt}
\begin{tabular}{|c|c|c|c|}
\hline
 Case     & SECL & Algorithm \ref{Pa1}&Algorithm \ref{Pa2}\\
\hline  1 &  0.3812 &0.3730&0.3761\\
\hline  2 &  0.5684 &0.5696&0.5678\\
\hline  3 &  0.2798 &0.2750&0.2712\\
\hline
\end{tabular}
\end{center}\vspace{-15pt}
\end{table}
\subsection{Comparison with SEHI\label{sectionNEb}}
\begin{figure}
\vspace{-15pt}
  \begin{center}
\includegraphics[ width=3.4in]{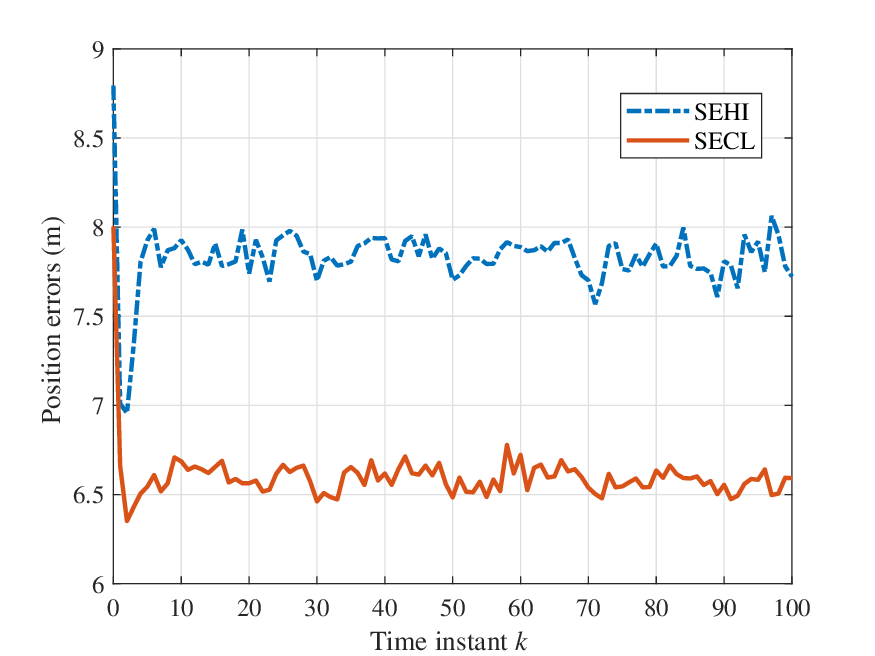}
\caption{  RMS Position errors of SEHI and SECL under the same average communication rate 0.35.}
  \label{figCp}
  \end{center}
\end{figure}

\begin{figure}
\vspace{-15pt}
  \begin{center}
\includegraphics[ width=3.4in]{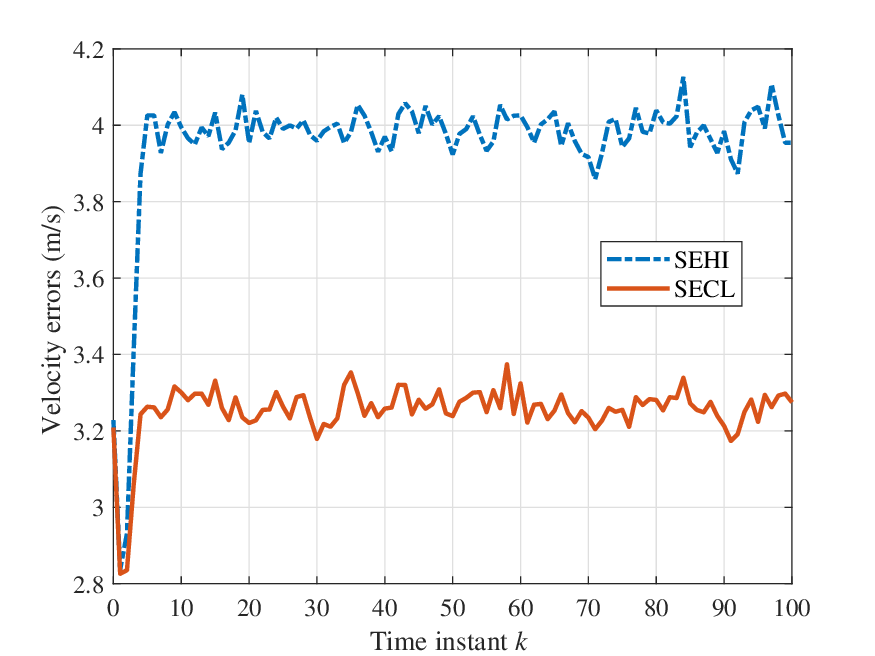}
\caption{    RMS velocity errors of SEHI and SECL under the same average communication rate 0.35.}
  \label{figCv}\vspace{-15pt}
  \end{center}
\end{figure}

We compare SECL with SEHI still using the above target tracking example where we take the same parameter values except for the tolerable upper bound $\overline{N}$.
In the comparison\vspace{2pt} with SEHI, we take\vspace{2pt} $\overline{N}=\overline{N}_a \triangleq 0.695\times\left(
                        \begin{array}{cc}
                          60 & 10\\
                          10 & 20 \\
                        \end{array}
                      \right)
$ which is a slight change of $\overline{N}$ at Case 3 in  Section \ref{sectionNEa} through multiplying by 0.695. After a Monte Carlo simulation, we get the average communication rate of SECL is $0.35$ when $\overline{N}=\overline{N}_a$. Considering fairness, we compare SECL with SEHI under the same average communication
rate. Using Monte Carlo simulation, we get $\delta=1.5565$ when the average communication rate of SEHI is $0.35$.
The target's RMS position errors of SECL and SEHI under the same average communication rate 0.35 are provided in Fig. \ref{figCp}, and the corresponding velocity errors are provided in Fig. \ref{figCv}.
From observing Figs. \ref{figCp} and \ref{figCv}, we find that SECL performs better than SEHI in target tracking accuracy for both position and velocity.
Hence, the simulation results show that SECL yields better tracking performance in contrast to SEHI.
\section{Conclusion \label{sectionCo}}
Based on confidence level, a novel event-triggered scheme has been proposed using the chi-square
distribution and regular Gaussian assumption. The novel scheme was applied to the state estimation problem for discrete-time linear systems in the environment of wireless sensor network so that a MMSE state estimator was proposed.
Two algorithms for estimating the communication rate of the proposed state estimator have been developed where, at time step $k$,  the first algorithm
is based on information up to $k-1$, and the second algorithm
is based on information up to $k-2$. A target tracking scenario has been given to examine the performance of the proposed results, and the simulation results
have shown that the proposed state estimator performs better than SEHI under the same average communication
rate. The simulation results have also shown that Algorithm \ref{Pa2} provides a good estimate for the communication rate of the proposed state estimator.

 \appendices
\section{Proof of Lemmas \ref{lemfGh}$-$\ref{lemfOm} \label{appenT1}}
\textit{Proof of Lemma \ref{lemfGh}:} Using  (\ref{sxk}), we have
\newpage
\noindent
\begin{align}
f(x_{k}|\textrm{I}_{k-1})=f(Ax_{k-1}+\omega_{k-1}|\textrm{I}_{k-1}).\label{PlemfGha}
\end{align}
From Assumptions 1 and 2, we see that $\omega_{k-1}$ is independent of $x_{k-1}$ and $\textrm{I}_{k-1}$. Hence,
$f(x_{k}|\textrm{I}_{k-1})$ is a linear combination of two Gaussian and mutually independent random vectors  $f(x_{k-1}|\textrm{I}_{k-1})$ and $\omega_{k-1}$, that is
$Af(x_{k-1}|\textrm{I}_{k-1})+\omega_{k-1}$. Then, using Remark \ref{Remarkeig}, we prove 1).
Applying (\ref{syk}), we have
$f(y_{k}|\textrm{I}_{k-1})=f(Cx_{k}+\upsilon_{k}|\textrm{I}_{k-1})$.
Then, using 1) of Lemma  \ref{lemfGh}, and referring to the proof of 1) of Lemma  \ref{lemfGh}, we prove 2).
\par
\textit{Proof of Lemma \ref{lemfGha}:}
1). Using  (\ref{zk}), (\ref{notb}) and (\ref{syk}), we get
\begin{align}
z_k=\varPhi Cx_k+\varPhi \upsilon_k-\varPhi\hat{y}_{k,k-1}.\label{PlemfGhb}
\end{align}
Then, using 1) of Lemma \ref{lemfGh}, and making reference to the proof of 1) of Lemma  \ref{lemfGh}, we prove 1).
\\
2). Applying (\ref{zk}), (\ref{notb}) and (\ref{nota}),
we get
\begin{align}
\textrm{E}[z_k|\textrm{I}_{k-1}]=\varPhi\textrm{E}[\tilde{y}_k|\textrm{I}_{k-1}]=\varPhi(\hat{y}_{k,k-1}-\hat{y}_{k,k-1})
=0.\label{og0l}
\end{align}
Then, applying 1) of Lemma \ref{lemfGha}, we prove 2).
\par
\textit{Proof of Lemma \ref{lemfOm}:}
1). From  (\ref{et}) and (\ref{ySya}), it follows that $\varphi_k\leq \chi_\alpha^2(p)$ is equivalent to $\sum_{i=1}^pz_{k,i}^2\leq \chi_\alpha^2(p)$.
Then, using the definition of $\Omega_k$ presented in  (\ref{Om}), we obtain
\begin{align}
\Omega_k= \bigg\{z_{k,1},z_{k,2},\cdots,z_{k,p}\in\mathbb{R}\bigg|\sum_{i=1}^pz_{k,i}^2\leq \chi_\alpha^2(p)\bigg\}.\label{PlfOma}
\end{align}
Then, using  (\ref{PlfOma}) and 2) of Lemma \ref{lemfGha}, we prove 1).\\
2).
Noticing \(z_{k}\in\mathbb{R}^{p}\), we have $z_kz_k^\textrm{T}=(\zeta_{k,ij})_{p\times p}$ with $\zeta_{k,ij}=z_{k,i}z_{k,j}$.
Then, using 2) of Lemma \ref{lemfGha}, we have
\begin{align}
\int_{\Omega_k}&f(z_k|\textrm{I}_{k-1})z_kz_k^\textrm{T} dz_k\nonumber\\
=&\int_{\Omega_k}\frac{g(z_k)}{(2\pi)^{0.5p}\big|N_k^{[z]}\big|^{0.5}}(\zeta_{k,ij})_{p\times p}dz_{k,1}dz_{k,2}\cdots dz_{k,p}\nonumber\\
=&\frac{1}{(2\pi)^{0.5p}\big|N_k^{[z]}\big|^{0.5}}\bigg(\int_{\Omega_k}g(z_k)\zeta_{k,ij}dz_{k,1}dz_{k,2}\cdots dz_{k,p}\bigg)_{\!\!\!p\times p}\label{PlfOmb}.
\end{align}
Using (\ref{PlfOma}) and noticing  $\zeta_{k,ij}=z_{k,i}z_{k,j}$, we have
\begin{align}
\int_{\Omega_k}&g(z_k)\zeta_{k,ij}dz_{k,1}dz_{k,2}\cdots dz_{k,p}\nonumber\\
&=\int\limits_{-\check{b}}^{\check{b}}\!\! dz_{k,1}\!\!\!\!
\int\limits_{-\check{z}_{k,1}}^{\check{z}_{k,1}}\!\!\!\! dz_{k,2}\cdots\!\!\!\!\!\! \int\limits_{-\check{z}_{k,p-1}}^{\check{z}_{k,p-1}}g(z_k)z_{k,i}z_{k,j}dz_{k,p}
\label{PlfOmc}
\end{align}
where $\check{b}$ and $\check{z}_{k,i}$ with $i=1,2,\cdots,p-1$ are defined in 1) of Lemma \ref{lemfOm}. Substituting (\ref{PlfOmc}) into (\ref{PlfOmb}), as well as using the definition
of $\psi_{k,ij}$ and $\varPsi_{k}$ given in 2) of Lemma \ref{lemfOm}, we have
\begin{align}
\int_{\Omega_k}f(z_k|\textrm{I}_{k-1})z_kz_k^\textrm{T} dz_k
=&\frac{(\psi_{k,ij})_{p\times p}}{(2\pi)^{0.5p}\big|N_k^{[z]}\big|^{0.5}}=\frac{\varPsi_{k}}{(2\pi)^{0.5p}\big|N_k^{[z]}\big|^{0.5}}.\label{PlfOmd}
\end{align}
This complete the proof of the statement.\vspace{-2pt}
\section{Proof of  Theorem  \ref{Theorem1}  \label{appenT3}}
Derivation of  (\ref{The1a}).    When  $\gamma_k=0$,
we have\vspace{-0.1pt}
\begin{align}
\hat{x}_k=\int_{\mathbb{R}^n}x_kf(x_k|\textrm{I}_{k})dx_k=\int_{\mathbb{R}^n}x_kf(x_k|\textrm{I}_{k-1},\gamma_k=0)dx_k.\label{og0a}
\end{align}
By (\ref{et}),
$f(x_k|\textrm{I}_{k-1},\gamma_k=0)$ can be expressed as
\begin{align}
f(x_k|\textrm{I}_{k-1},\gamma_k=0)=&f\big(x_k|\textrm{I}_{k-1},\varphi_k\leq \chi_\alpha^2(p)\big)\nonumber\\
=&f(x_k|\textrm{I}_{k-1},z_k\in\Omega_k)\nonumber\\
=&\frac{f(x_k,\textrm{I}_{k-1},z_k\in\Omega_k)}{f(\textrm{I}_{k-1},z_k\in\Omega_k)}\nonumber\\
=&\frac{\int_{\Omega_k}f(x_k,\textrm{I}_{k-1},z_k)dz_k}{\int_{\Omega_k}f(\textrm{I}_{k-1},z_k)dz_k}\nonumber\\
=&\frac{\int_{\Omega_k}f(x_k|\textrm{I}_{k-1},z_k)f(z_k|\textrm{I}_{k-1})dz_k}{\int_{\Omega_k}f(z_k|\textrm{I}_{k-1})dz_k}\label{og0b}
\end{align}
where the second and last equalities are due to (\ref{Om}) and  Bayes' rule, respectively.
Substituting (\ref{og0b}) into (\ref{og0a}) yields that
\begin{align}
\hat{x}_k=&\int_{\mathbb{R}^n}x_k\frac{\int_{\Omega_k}f(x_k|\textrm{I}_{k-1},z_k)f(z_k|\textrm{I}_{k-1})dz_k}{\int_{\Omega_k}f(z_k|\textrm{I}_{k-1})dz_k}dx_k\nonumber\\
=&\frac{\int_{\mathbb{R}^n}x_k\int_{\Omega_k}f(x_k|\textrm{I}_{k-1},z_k)f(z_k|\textrm{I}_{k-1})dz_k dx_k}{\int_{\Omega_k}f(z_k|\textrm{I}_{k-1})dz_k}\nonumber\\
=&\frac{\int_{\Omega_k}f(z_k|\textrm{I}_{k-1})\int_{\mathbb{R}^n}x_kf(x_k|\textrm{I}_{k-1},z_k)dx_k dz_k }{\int_{\Omega_k}f(z_k|\textrm{I}_{k-1})dz_k}\nonumber\\
=&\frac{\int_{\Omega_k}f(z_k|\textrm{I}_{k-1})\hat{x}_k^{[z]} dz_k }{\int_{\Omega_k}f(z_k|\textrm{I}_{k-1})dz_k}
.\label{og0c}
\end{align}
Making reference to the proof of Theorem 3.2 in \cite{et1}, we
can obtain
\begin{align}
\frac{\int_{\Omega_k}f(z_k|\textrm{I}_{k-1})\hat{x}_k^{[z]} dz_k }{\int_{\Omega_k}f(z_k|\textrm{I}_{k-1})dz_k}=&\hat{x}_{k,k-1}.\label{og0d}
\end{align}
where $\hat{x}_{k,k-1}$ is defined in (\ref{LeGhna}).
Substituting  (\ref{og0d}) into  (\ref{og0c}), we derive (\ref{The1a}).
\\
Derivation of (\ref{The1d}).
Using (\ref{og0k}) and the definition of conditional covariance matrix,
we have
\begin{align}
N_k^{[z]}= \textrm{E}\big[(z_k-\textrm{E}[z_k|\textrm{I}_{k-1}])(z_k-\textrm{E}[z_k|\textrm{I}_{k-1}])^\textrm{T}\big].\label{og0mA}
\end{align}
Substituting (\ref{og0l}) into (\ref{og0mA}), and using (\ref{zk}), we derive
\begin{align}
N_k^{[z]}= \textrm{E}\big[z_kz_k^\textrm{T}\big]=\textrm{E}\big[\varPhi\tilde{y}_k\tilde{y}_k^\textrm{T}\varPhi^\textrm{T}\big]
=\varPhi(CM_kC^\textrm{T}+R)\varPhi^\textrm{T}\label{og0m}
\end{align}
where $M_k$ is defined in  (\ref{LeGhnb}).
Similarly, we have
\begin{align}
\textrm{E}\Big[\tilde{x}_k(z_k-[z_k|\textrm{I}_{k-1}])^\textrm{T}\Big]=\textrm{E}\Big[\tilde{x}_kz_k^\textrm{T}\Big]=M_k C^\textrm{T}\varPhi^\textrm{T}.
\label{og0n}
\end{align}
Substituting  (\ref{og0m}) and (\ref{og0n}) into  (\ref{og0j}) yields (\ref{The1d}).
\\
Derivation of (\ref{The1c}).
Utilizing Kalman filter, we get
\begin{align}
\hat{x}_k^{[z]}=\hat{x}_{k,k-1}+K_k(z_k-\textrm{E}[z_k|\textrm{I}_{k-1}])\label{og0iA}
\end{align}
where $\hat{x}_k^{[z]}$ and $K_k$ are defined in  (\ref{notc}) and  (\ref{og0j}), respectively, and $N_k^{[z]}$ in $K_k$ is defined in (\ref{og0k}).\vspace{-0.1pt}
Substituting  (\ref{og0l}) into  (\ref{og0iA}), we obtain
\begin{align}
\hat{x}_k^{[z]}=\hat{x}_{k,k-1}+K_kz_k.\label{og0i}
\end{align}
From  (\ref{notc1}),  (\ref{notc}) and the definition of conditional covariance matrix, it follows that
\begin{align}
P_k^{[z]}=&\textrm{E}\Big[\big(x_k-\hat{x}_k^{[z]}\big)\big(x_k-\hat{x}_k^{[z]}\big)^\textrm{T}\Big].
\label{og0iB}
\end{align}
Substituting  (\ref{og0i}) into (\ref{og0iB}) and using (\ref{notb2}), we obtain\vspace{-0.1pt}
\begin{align}
P_k^{[z]}=&\textrm{E}\big[(\tilde{x}_k-K_kz_k)(\tilde{x}_k-K_kz_k)^\textrm{T}\big]\nonumber\\
=&M_k-M_k C^\textrm{T}\varPhi^\textrm{T}K_k^\textrm{T}-K_k\varPhi CM_k+K_kN_k^{[z]} K_k^\textrm{T}\label{og0ia}
\end{align}
where $M_k$ is defined in (\ref{LeGhnb}), and the last equality is due to (\ref{og0m}) and (\ref{og0n}).
Applying (\ref{The1d}) and (\ref{og0m}), we easily obtain
\begin{align}
M_k C^\textrm{T}\varPhi^\textrm{T}K_k^\textrm{T}=&K_k\varPhi CM_k\nonumber\\
=&K_kN_k^{[z]} K_k^\textrm{T}=M_kC^\textrm{T}(CM_kC^\textrm{T}+R)^{-1}CM_k.\label{og0ia1}
\end{align}
Substituting (\ref{og0ia1}) into (\ref{og0ia}), we obtain (\ref{The1c}).
\\
Derivation of (\ref{The1P}).
Using (\ref{notb1}) and (\ref{The1a}), we have
\begin{align}
P_k=&\textrm{E}\big[(x_k-\hat{x}_k)(x_k-\hat{x}_k)^\textrm{T}\big]\nonumber\\
=&\int_{\mathbb{R}^n}(x_k-\hat{x}_k)(x_k-\hat{x}_k)^\textrm{T}f(x_k|\textrm{I}_{k-1},\gamma_k=0)dx_k\nonumber\\
=&\int_{\mathbb{R}^n}\tilde{x}_k\tilde{x}_k^\textrm{T}f(x_k|\textrm{I}_{k-1},\gamma_k=0)dx_k
\label{og0f}
\end{align}
where $\tilde{x}_k$ is defined in (\ref{notb2}).
Substituting (\ref{og0b}) into (\ref{og0f}), and referring to (\ref{og0c}), we obtain
\begin{align}
P_k=&\frac{\int_{\Omega_k}f(z_k|\textrm{I}_{k-1})M_k^{[z]} dz_k }{\int_{\Omega_k}f(z_k|\textrm{I}_{k-1})dz_k}
\label{og0g}
\end{align}
where
\begin{align}
M_k^{[z]}\triangleq \int_{\mathbb{R}^n}\tilde{x}_k\tilde{x}_k^\textrm{T}f(x_k|\textrm{I}_{k-1},z_k)dx_k.\label{og0h}
\end{align}
Using (\ref{og0h}) and making reference to the proof of Theorem 3.2 in \cite{et1}, we obtain\vspace{-0.1pt}
\begin{align}
M&_k^{[z]}\nonumber\\
=& \int_{\mathbb{R}^n}(\tilde{x}_k-K_kz_k+K_kz_k)(\tilde{x}_k-K_kz_k+K_kz_k)^\textrm{T}f(x_k|\textrm{I}_{k-1},z_k)dx_k\nonumber\\
=&\bigg(\int_{\mathbb{R}^n}(\tilde{x}_k-K_kz_k)(\tilde{x}_k-K_kz_k)^\textrm{T}f(x_k|\textrm{I}_{k-1},z_k)dx_k\bigg)+K_kz_kz_k^\textrm{T}K_k^\textrm{T}\nonumber\\
=&P_k^{[z]}+K_kz_kz_k^\textrm{T}K_k^\textrm{T}.\label{og0q}
\end{align}
Substituting  (\ref{og0q}) into (\ref{og0g}), we get
\begin{align}
P_k=&\frac{\int_{\Omega_k}f(z_k|\textrm{I}_{k-1})(P_k^{[z]}+K_kz_kz_k^\textrm{T}K_k^\textrm{T}) dz_k }{\int_{\Omega_k}f(z_k|\textrm{I}_{k-1})dz_k}\nonumber\\
=&P_k^{[z]}+\frac{K_k\int_{\Omega_k}f(z_k|\textrm{I}_{k-1})z_kz_k^\textrm{T} dz_k K_k^\textrm{T}}{\int_{\Omega_k}f(z_k|\textrm{I}_{k-1})dz_k}\nonumber\\
=&P_k^{[z]}+\frac{1}{h_k}K_k\varPsi_{k} K_k^\textrm{T}
\label{og0r}
\end{align}
where the last equality is because of Lemma \ref{lemfOm}. This completes the derivation of (\ref{The1P}).
\section{Proof of  Theorem \ref{Theorem2} \label{appenT4}}
The following lemma is required to obtain Theorem \ref{Theorem2}.
\begin{lemma}
Under the assumption that $f(x_{k-2}|\textrm{I}_{k-2})$ is Gaussian, it holds that
 $\textrm{E}[y_{k}|\textrm{I}_{k-2},y_{k-1}]=\textrm{E}[y_{k}|\textrm{I}_{k-2},z_{k-1}]$.
\label{lemyeqz}
\end{lemma}
\textit{Proof:}
$f(y_{k}|\textrm{I}_{k-2},z_{k-1})$ can be given by
\begin{align}
f(y_{k}|\textrm{I}_{k-2},z_{k-1})=&\frac{f(\textrm{I}_{k-2},z_{k-1},y_{k})}{f(\textrm{I}_{k-2},z_{k-1})}\nonumber\\
=&\frac{f(z_{k-1},y_{k}|\textrm{I}_{k-2})}{f(z_{k-1}|\textrm{I}_{k-2})}
=\frac{f(\eta_{k}|\textrm{I}_{k-2})}{f(z_{k-1}|\textrm{I}_{k-2})}
\label{Plemyeqz0}
\end{align}
where $\eta_{k}\triangleq (z_{k-1}^\textrm{T},y_{k}^\textrm{T})^\textrm{T}$. Under the assumption that $f(x_{k-2}|\textrm{I}_{k-2})$ is Gaussian, we easily see that $f(\eta_{k}|\textrm{I}_{k-2})$
is Gaussian by making reference to the proof of Lemmas \ref{lemfGh} and \ref{lemfGha}.
Hence, we have
\begin{align}
f(\eta_{k}|\textrm{I}_{k-2})=\frac{g_1({\eta}_{k})}{(2\pi)^{0.5\times2p}\big|N_k^{[\eta]}\big|^{0.5}}=\frac{g_1({\eta}_{k})}{(2\pi)^{p}\big|N_k^{[\eta]}\big|^{0.5}}
\label{Plemyeqz1}
\end{align}
where
\begin{align}
g_1({\eta}_{k})\triangleq &\textrm{exp}\big\{\!\!\!-0.5(\eta_{k}-\hat{\eta}_{k,k-2})^\textrm{T}(N_k^{[\eta]})^{-1}(\eta_{k}-\hat{\eta}_{k,k-2})\big\},
\label{Plemyeqz1a}\\
\hat{\eta}_{k,k-2}\triangleq&\textrm{E}[\eta_{k}|\textrm{I}_{k-2}],N_k^{[\eta]}\triangleq \textrm{Var}(\eta_{k}|\textrm{I}_{k-2}).\label{Plemyeqz1b}
\end{align}
Utilizing Lemma \ref{lemfGha}, we have\vspace{-0.1pt}
\begin{align}
f(z_{k-1}|\textrm{I}_{k-2})=\frac{g(z_{k-1})}{(2\pi)^{0.5p}\big|N_{k-1}^{[z]}\big|^{0.5}}.\label{Plemyeqz1ba}
\end{align}
Substituting (\ref{Plemyeqz1}) and (\ref{Plemyeqz1ba}) into (\ref{Plemyeqz0}) yields that
\begin{align}
f(y_{k}|\textrm{I}_{k-2},z_{k-1})=\frac{g_1({\eta}_{k})\big|N_{k-1}^{[z]}\big|^{0.5}}{(2\pi)^{0.5p}g(z_{k-1})\big|N_k^{[\eta]}\big|^{0.5}}.
\label{Plemyeqz1bb}
\end{align}
In the same way, we can derive
\begin{align}
f(y_{k}|\textrm{I}_{k-2},y_{k-1})=\frac{g_1({\zeta}_{k})\big|N_{k-1}\big|^{0.5}}{(2\pi)^{0.5p}g_2(y_{k-1})\big|N_k^{[\zeta]}\big|^{0.5}}
\label{Plemyeqz1bb1}
\end{align}
where
\begin{align}
&g_1({\zeta}_{k})\triangleq\textrm{exp}\big\{\!\!\!-0.5(\zeta_{k}-\hat{\zeta}_{k,k-2})^\textrm{T}(N_k^{[\zeta]})^{-1}(\zeta_{k}-\hat{\zeta}_{k,k-2})\big\},
\label{Plemyeqz1bc}\\
&\zeta_{k}\triangleq (y_{k-1}^\textrm{T},y_{k}^\textrm{T})^\textrm{T},\hat{\zeta}_{k,k-2}\triangleq\textrm{E}[\zeta_{k}|\textrm{I}_{k-2}],N_k^{[\zeta]}\triangleq \textrm{Var}(\zeta_{k}|\textrm{I}_{k-2}),\label{Plemyeqz1bd}\\
&g_2(y_{k-1})\triangleq \textrm{exp}\big\{\!\!\!-0.5\tilde{y}_{k-1}^\textrm{T}N_{k-1}^{-1}\tilde{y}_{k-1}\big\}.\label{Plemyeqz1be}
\end{align}
Using (\ref{zk}) and (\ref{og0l}), we have
\begin{align}
\eta_{k}=\left(
                  \begin{array}{c}
                    \varPhi\tilde{y}_{k-1} \\
                    y_k \\
                  \end{array}
                \right), \hat{\eta}_{k,k-2}=\left(
                                              \begin{array}{c}
                                                0 \\
                                               \textrm{E}[y_{k}|\textrm{I}_{k-2}] \\
                                              \end{array}
                                            \right)
\label{Plemyeqz2}
\end{align}
which means
\begin{align}
\eta_{k}-\hat{\eta}_{k,k-2}=\left(
                  \begin{array}{c}
                    \varPhi\tilde{y}_{k-1} \\
                    y_k-\textrm{E}[y_{k}|\textrm{I}_{k-2}] \\
                  \end{array}
                \right)=\varPhi_a(\zeta_{k}-\hat{\zeta}_{k,k-2})\label{Plemyeqz3}
\end{align}
where\vspace{-0.1pt}
\begin{align}
\varPhi_a\triangleq\left(
            \begin{array}{cc}
              \varPhi & 0 \\
              0 & \textsl{\textsf{I}}_p \\
            \end{array}
          \right).
\label{Plemyeqz4}
\end{align}
Using the  definition of conditional covariance matrix and utilizing (\ref{Plemyeqz3}),
we easily obtain
\begin{align}
N_k^{[\eta]}=\varPhi_aN_k^{[\zeta]}\varPhi_a^\textrm{T}
\label{Plemyeqz5}
\end{align}
where $N_k^{[\eta]}$ and $N_k^{[\zeta]}$ are defined in (\ref{Plemyeqz1b}) and (\ref{Plemyeqz1bd}), respectively.\vspace{-0.1pt}
Substituting  (\ref{Plemyeqz3}) and (\ref{Plemyeqz5}) into  (\ref{Plemyeqz1a}) yields that
\begin{align}
g_1({\eta}_{k})= &\textrm{exp}\big\{\!\!\!-0.5(\zeta_{k}-\hat{\zeta}_{k,k-2})^\textrm{T}\varPhi_a^\textrm{T}\varPhi_a^{-\textrm{T}}(N_k^{[\zeta]})^{-1}\varPhi_a^{-1}\varPhi_a(\zeta_{k}\nonumber\\
&-\hat{\zeta}_{k,k-2})\big\}\nonumber\\
=&\textrm{exp}\big\{\!\!\!-0.5(\zeta_{k}-\hat{\zeta}_{k,k-2})^\textrm{T}(N_k^{[\zeta]})^{-1}(\zeta_{k}-\hat{\zeta}_{k,k-2})\big\}.
\label{Plemyeqz6}
\end{align}
Putting (\ref{Plemyeqz6}) and (\ref{Plemyeqz1bc}) together, we have\vspace{-0.1pt}
\begin{align}
g_1({\eta}_{k})=g_1({\zeta}_{k}).
\label{Plemyeqz7}
\end{align}
In the same way, we can obtain
\begin{align}
g(z_{k-1})=g_2(y_{k-1}).
\label{Plemyeqz8}
\end{align}
From (\ref{Plemyeqz5}), it follows that
\begin{align}
\big|N_k^{[\eta]}\big|=\big|\varPhi_a\big|\big|N_k^{[\zeta]}\big|\big|\varPhi_a^\textrm{T}\big|=\big|\varPhi_a\big|^2\big|N_k^{[\zeta]}\big|
\label{Plemyeqz9}
\end{align}
where the last equality is because the determinant of a matrix is equal to the determinant of its transpose.
Noting that $\varPhi_a$ defined in (\ref{Plemyeqz4}) is a block-diagonal matrix, we have
\begin{align}
|\varPhi_a|=|\varPhi||\textsl{\textsf{I}}_p|=|\varPhi|.
\label{Plemyeqz10}
\end{align}
Substituting (\ref{Plemyeqz10}) into (\ref{Plemyeqz9}) yields that
\begin{align}
\big|N_k^{[\eta]}\big|=\big|\varPhi\big|^2\big|N_k^{[\zeta]}\big|.
\label{Plemyeqz11}
\end{align}
Similarly, we can get
\begin{align}
\big|N_{k-1}^{[z]}\big|=|\varPhi|^2|N_{k-1}|.
\label{Plemyeqz12}
\end{align}
Substituting (\ref{Plemyeqz7}), (\ref{Plemyeqz8}), (\ref{Plemyeqz11}) and (\ref{Plemyeqz12}) into (\ref{Plemyeqz1bb}), we derive
\begin{align}
f(y_{k}|\textrm{I}_{k-2},z_{k-1})=&\frac{g_1({\zeta}_{k})(|\varPhi|^2|N_{k-1}|)^{0.5}}{(2\pi)^{0.5p}g_2(y_{k-1})\big(\big|\varPhi\big|^2\big|N_k^{[\zeta]}\big|\big)^{0.5}}\nonumber\\
=&\frac{g_1({\zeta}_{k})|N_{k-1}|^{0.5}}{(2\pi)^{0.5p}g_2(y_{k-1})\big|N_k^{[\zeta]}\big|^{0.5}}.\label{Plemyeqz13}
\end{align}
Combining (\ref{Plemyeqz1bb1}) and (\ref{Plemyeqz13}), we derive $f(y_{k}|\textrm{I}_{k-2},y_{k-1})=f(y_{k}|\textrm{I}_{k-2},z_{k-1})$ which means Lemma \ref{lemyeqz} holds.
\par
Now, we provide a proof of Theorem \ref{Theorem2}.
\\
 Derivation of  (\ref{The2b}).
Substituting (\ref{sxk}) at time step $k$  into  (\ref{syk}), we obtain\vspace{-1pt}
\begin{align}
y_k=CAx_{k-1}+C\omega_{k-1}+\upsilon_{k}.\label{Ega61}
\end{align}
When $\gamma_{k-1}=1$, we have  $\textrm{I}_{k-1}=(\textrm{I}_{k-2},y_{k-1})$.
Hence, we have
\begin{align}
\hat{y}_{k,k-1}=\textrm{E}[y_k|\textrm{I}_{k-2},y_{k-1}]=\textrm{E}[y_{k}|\textrm{I}_{k-2},z_{k-1}]\label{Ega2aA1}
\end{align}
where $\hat{y}_{k,k-1}$ is defined in (\ref{nota}), and the last equality is due to Lemma \ref{lemyeqz}.
Substituting (\ref{zk}) into (\ref{Nca0}), we get\vspace{-0.1pt}
\begin{align}
\hat{z}_{k,k-1}^\triangleright=\varPhi\textrm{E}[\tilde{y}_k|\textrm{I}_{k-2},z_{k-1}]=0\label{Ega2aA2}
\end{align}
where $\tilde{y}_k$ is defined in (\ref{notb}), and the last equality is due to (\ref{Ega2aA1}).
Substituting (\ref{Ega2aA2}) into (\ref{Nca5}), we derive
\begin{align}
\tilde{z}_{k}^\triangleright= &z_{k}.\label{The2a}
\end{align}
Substituting (\ref{Ega61}) into (\ref{nota}) at $\textrm{I}_{k-1}=(\textrm{I}_{k-2},y_{k-1})$, as well as using Assumptions 1 and 2, we get
\begin{align}
\hat{y}_{k,k-1}=&CA\textrm{E}[{x}_{k-1}|\textrm{I}_{k-2},y_{k-1}]\nonumber\\
=&CA(\hat{x}_{k-1,k-2}+K_{k-1}\varPhi\tilde{y}_{k-1})=CA\hat{x}_{k-1}^{[z]}\label{Ega4}
\end{align}
where the second equality is due to the Kalman filter and (\ref{Al1e}), and the last equality is because of (\ref{zk}) and (\ref{og0i}).
Substituting  (\ref{sxk}) and (\ref{Ega4}) into  (\ref{PlemfGhb}), we obtain
\begin{align}
z_k=&\varPhi CAx_{k-1}+\varPhi C\omega_{k-1}+\varPhi \upsilon_k-\varPhi CA\hat{x}_{k-1}^{[z]}\nonumber
\end{align}
\begin{align}
=&\varPhi \big(CA(x_{k-1}-\hat{x}_{k-1}^{[z]})+C\omega_{k-1}+\upsilon_k\big)\label{Ega4a}
\end{align}
given $(\textrm{I}_{k-2},z_{k-1})$.
Utilizing (\ref{Nca0})$-$(\ref{Nca1}) and the definition of conditional covariance matrix, we have
\begin{align}
\vec{N}_k^{[z]}=\textrm{E}\big[\tilde{z}_{k}^{\triangleright}(\tilde{z}_{k}^{\triangleright})^\textrm{T}\big].
\label{EgaA5}
\end{align}
Substituting (\ref{The2a}) into
 (\ref{EgaA5}) and using (\ref{Ega4a}), we get\vspace{-0.1pt}
\begin{align}
\vec{N}_k^{[z]}=&\textrm{E}\Big[\big(\varPhi \big(CA(x_{k-1}-\hat{x}_{k-1}^{[z]})+C\omega_{k-1}+\upsilon_k\big)\big)\big(\varPhi \big(CA(x_{k-1}\nonumber\\
&-\hat{x}_{k-1}^{[z]})+C\omega_{k-1}+\upsilon_k\big)\big)^\textrm{T}\Big]\nonumber\\
=&\varPhi\big(C(AP_{k-1}^{[z]} A^\textrm{T}+Q) C^\textrm{T}+R)\varPhi^\textrm{T}\label{Ega5}
\end{align}
where $P_{k}^{[z]}$ is defined in (\ref{notc1}).
This completes the derivation of  (\ref{The2b}).
\\
 Derivation of  (\ref{The2b1}).
 Making reference to (\ref{Ega2aA1}) and (\ref{Ega2aA2}), we get
\begin{align}
\hat{z}_{k,k-1}(0)=\varPhi\textrm{E}[\tilde{y}_k|\textrm{I}_{k-2},\gamma_{k-1}=0]=0\label{Ega2aA21}
\end{align}
where $\hat{z}_{k,k-1}(0)$ is defined in (\ref{Nca6}).
Substituting (\ref{Ega61}) into (\ref{nota}), as well as using Assumptions 1 and 2, we have
\begin{align}
\hat{y}_{k,k-1}=&CA\hat{x}_{k-1}.\label{Ega2aC}
\end{align}
Substituting  (\ref{sxk}) and (\ref{Ega2aC}) into  (\ref{PlemfGhb}) yields that
\begin{align}
z_k=&\varPhi CAx_{k-1}+\varPhi C\omega_{k-1}+\varPhi \upsilon_k-\varPhi CA\hat{x}_{k-1}\nonumber\\
=&\varPhi \big(CA(x_{k-1}-\hat{x}_{k-1})+C\omega_{k-1}+\upsilon_k\big).\label{Ega2aD}
\end{align}
Using (\ref{Nca7}), (\ref{Nca6}) and the definition of conditional covariance matrix, we obtain
\begin{align}
N_k(0)= &\textrm{E}\Big[\big(z_k-\hat{z}_{k,k-1}(0)\big)\big(z_k-\hat{z}_{k,k-1}(0)\big)^\textrm{T}\Big].
\label{Ega2aDA1}
\end{align}
Substituting  (\ref{Ega2aD}) and  (\ref{Ega2aA21}) into (\ref{Ega2aDA1}), we get
\begin{align}
N_k(0)= &\textrm{E}\Big[\big(\varPhi (CA(x_{k-1}-\hat{x}_{k-1})+C\omega_{k-1}+\upsilon_k)\big)\big(\varPhi (CA(x_{k-1}\nonumber\\
&-\hat{x}_{k-1})+C\omega_{k-1}+\upsilon_k)\big)^\textrm{T}\Big]\nonumber\\
=&\varPhi\Big(C\big(A(P_{k-1}^{[z]}+\frac{1}{h_{k-1}}K_{k-1}\varPsi_{k-1} K_{k-1}^\textrm{T})A^\textrm{T}+Q\big)C^\textrm{T}+R\Big)\nonumber\\
&\times\varPhi^\textrm{T}\label{Ega2aD1}
\end{align}
where the last equality is due to (\ref{notb1}) and (\ref{The1P}) of Theorem \ref{Theorem1} for given $(\textrm{I}_{k-2},\gamma_{k-1}=0)$.
This completes the derivation of (\ref{The2b1}).
\\
 Derivation of  (\ref{The2c}) and  (\ref{The2ca}).
 When $f(x_{k-1}|\textrm{I}_{k-1})$ is Gaussian, we see that $f(x_{k-1}|\textrm{I}_{k-2})$ is Gaussian. Then, using
Remark \ref{Remarkeig} and (\ref{Ega4a}), we conclude that $f(z_{k}|\textrm{I}_{k-2},z_{k-1})$ is Gaussian.
Hence, we have
\begin{align}
f(z_{k}|\textrm{I}_{k-2},z_{k-1})=&\frac{\textrm{exp}\big\{\!\!\!-0.5(\tilde{z}_{k}^\triangleright)^\textrm{T}(\vec{N}_k^{[z]})^{-1}\tilde{z}_{k}^\triangleright\big\}}{(2\pi)^{0.5p}\big|\vec{N}_k^{[z]}\big|^{0.5}}\label{pThe2c}
\end{align}
where $\tilde{z}_{k}^\triangleright$ and $\vec{N}_k^{[z]}$ are defined in (\ref{Nca5}) and (\ref{Nca1}), respectively.
Substituting (\ref{The2a}) into (\ref{pThe2c}) and replacing $\textrm{exp}\big\{\!\!\!-0.5{z}_k^\textrm{T}(\vec{N}_k^{[z]})^{-1}{z}_k\big\}$ by $\vec{g}(z_k)$ yield (\ref{The2c}).
Similarly to the derivation of (\ref{The2c}), we can obtain  (\ref{The2ca}) by using (\ref{Ega2aA21}) and (\ref{Nca7}).
\\
 Derivation of  (\ref{The2d}) and (\ref{The2da}).
$P(\gamma_k=0|\textrm{I}_{k-2},z_{k-1})$ can be given by\vspace{-0.1pt}
\begin{align}
P(\gamma_k=0|\textrm{I}_{k-2},z_{k-1})=&P(z_{k}\in\Omega_{k}|\textrm{I}_{k-2},z_{k-1})\nonumber\\
=&\int_{\Omega_{k}}f(z_{k}|\textrm{I}_{k-2},z_{k-1})dz_{k}.
\label{Ega3}
\end{align} Then, replacing $P(\gamma_k=0|\textrm{I}_{k-2},z_{k-1})$ by $\vec{P}_k^{[z]}(0)$, we derive (\ref{The2d}). In the same way, we can obtain (\ref{The2da})
where $\breve{P}_k(0)$ is defined in (\ref{Nop1aA}).
\\
 Derivation of (\ref{The2e}).
$P(\gamma_k=0|\textrm{I}_{k-2})$ can be given by
\begin{align}
P(&\gamma_k=0|\textrm{I}_{k-2})\nonumber\\
=&P(\gamma_{k-1}=0|\textrm{I}_{k-2})P(\gamma_k=0|\textrm{I}_{k-2},\gamma_{k-1}=0)\nonumber\\
&+P(\gamma_{k-1}=1|\textrm{I}_{k-2})P(\gamma_k=0|\textrm{I}_{k-2},\gamma_{k-1}=1).
\label{Ega2}
\end{align}
$P(\gamma_k=0|\textrm{I}_{k-2},\gamma_{k-1}=1)$ can be given by
\begin{align}
P(\gamma_k=0|&\textrm{I}_{k-2},\gamma_{k-1}=1)\nonumber\\
=&\frac{P(\textrm{I}_{k-2},\gamma_{k-1}=1,\gamma_k=0)}{P(\textrm{I}_{k-2},\gamma_{k-1}=1)}\nonumber\\
=&\frac{P(\textrm{I}_{k-2},z_{k-1}\in\breve{\Omega}_{k-1},\gamma_k=0)}{P(\textrm{I}_{k-2},z_{k-1}\in\breve{\Omega}_{k-1})}\nonumber\\
=&\frac{\int_{\breve{\Omega}_{k-1}}f(\textrm{I}_{k-2},z_{k-1},\gamma_k=0)dz_{k-1}}{\int_{\breve{\Omega}_{k-1}}f(\textrm{I}_{k-2},z_{k-1})dz_{k-1}}\nonumber\\
=&\frac{\int_{\breve{\Omega}_{k-1}}\vec{P}_k^{[z]}(0)f(\textrm{I}_{k-2},z_{k-1})dz_{k-1}}{\int_{\breve{\Omega}_{k-1}}f(\textrm{I}_{k-2},z_{k-1})dz_{k-1}}\nonumber\\
=&\frac{\int_{\breve{\Omega}_{k-1}}\vec{P}_k^{[z]}(0)f(z_{k-1}|\textrm{I}_{k-2})dz_{k-1}}{\int_{\breve{\Omega}_{k-1}}f(z_{k-1}|\textrm{I}_{k-2})dz_{k-1}}\nonumber\\
=&\frac{\int_{\breve{\Omega}_{k-1}}\vec{P}_k^{[z]}(0)f(z_{k-1}|\textrm{I}_{k-2})dz_{k-1}}{P(\gamma_{k-1}=1|\textrm{I}_{k-2})}\label{Ega2a}
\end{align}
where $\breve{\Omega}_{k}$ and $\vec{P}_k^{[z]}(0)$ are defined in (\ref{Omc}) and (\ref{Nop1a}), respectively. Substituting
(\ref{Ega2a}) into  (\ref{Ega2}), as well as replacing $P(\gamma_{k-1}=0|\textrm{I}_{k-2})$ by
$P_{k-1,k-2}(0)$, we obtain\vspace{-0.1pt}
 \begin{align}
 P(\gamma_k=0|\textrm{I}_{k-2})=&P_{k-1,k-2}(0)P(\gamma_k=0|\textrm{I}_{k-2},\gamma_{k-1}=0)\nonumber\\
 &+
 \int_{\breve{\Omega}_{k-1}}\vec{P}_k^{[z]}(0)f(z_{k-1}|\textrm{I}_{k-2})dz_{k-1}.
 \label{Ega3c}
\end{align}
Then, replacing $P(\gamma_k=0|\textrm{I}_{k-2})$ and $P(\gamma_k=0|\textrm{I}_{k-2},\gamma_{k-1}=0)$ by ${P}_{k,k-2}(0)$ and
$\breve{P}_k(0)$, respectively, we derive (\ref{The2e}). Making reference to (\ref{Ega0}), we easily obtain (\ref{The2f}).
\section{Proof of  Algorithm \ref{Pa2}  \label{appenT5}}
Making reference to the proof of 1) of Lemma \ref{lemfOm}, we can obtain (\ref{Al2d}) by using (\ref{The2c}) and (\ref{The2d}).
Similarly, we obtain (\ref{Al2e}) by using (\ref{The2ca}) and (\ref{The2da}).
From (\ref{The2d}), (\ref{The2c}) and  (\ref{Nca4}), we see that
$\vec{P}_k^{[z]}(0)$ does not contain the random vector $z_{k-1}$. Hence, we have
\begin{align}
&\!\!\!\!\!\!\!\!\!\!\!\!\int_{\breve{\Omega}_{k-1}}\vec{P}_k^{[z]}(0)f(z_{k-1}|\textrm{I}_{k-2})dz_{k-1}\nonumber\\
=&\vec{P}_k^{[z]}(0)\int_{\breve{\Omega}_{k-1}}f(z_{k-1}|\textrm{I}_{k-2})dz_{k-1}\nonumber\\
=&\vec{P}_k^{[z]}(0)\Big(\int_{\mathbb{R}^p}f(z_{k-1}|\textrm{I}_{k-2})dz_{k-1}-\int_{{\Omega}_{k-1}}f(z_{k-1}|\textrm{I}_{k-2})dz_{k-1}\Big)\nonumber
\end{align}
\begin{align}
=&\vec{P}_k^{[z]}(0)\Big(1-\int_{{\Omega}_{k-1}}f(z_{k-1}|\textrm{I}_{k-2})dz_{k-1}\Big)\nonumber\\
=&\vec{P}_k^{[z]}(0)\Big(1-\frac{h_{k-1}}{(2\pi)^{0.5p}\big|N_{k-1}^{[z]}\big|^{0.5}}\Big)\nonumber\\
=&\vec{P}_k^{[z]}(0)\big(1-P_{k-1,k-2}(0)\big)\label{PAl2e1}
\end{align}
where the fourth equality is due to  1) of Lemma \ref{lemfOm}, and the last equality is because of (\ref{Alo4a}).
Substituting (\ref{PAl2e1}) into  (\ref{The2e}), we get
\begin{align}
{P}_{k,k-2}(0)=&P_{k-1,k-2}(0)\breve{P}_k(0)
 +\vec{P}_k^{[z]}(0)
\big(1-P_{k-1,k-2}(0)\big)\nonumber\\
=&\vec{P}_k^{[z]}(0)+P_{k-1,k-2}(0)\big(\breve{P}_k(0)-\vec{P}_k^{[z]}(0)\big),\label{PAl2e2}
\end{align}
which means that (\ref{Al2f}) holds.

\end{document}